\documentclass{aa}  

\usepackage{multirow}
\usepackage{graphicx}
\usepackage{amsmath}

\usepackage[colorlinks=true,citecolor=blue, urlcolor=blue, linkcolor=blue]{hyperref}
\usepackage[varg]{txfonts}
\usepackage{upgreek} % for upmu

\makeatletter

\let\linenumbers\relax

\let\runninglinenumbers\relax

\let\runningpagewiselinenumbers\relax

\let\modulolinenumbers\relax

\makeatother

\begin{document}

%%%%%%%%%%%%%%%%%%%%%%%%%%%%%%%%%%%%%%%%
% if you use custom commands in your title,
% ensure to check your title when submitting!
%%%%%%%%%%%%%%%%%%%%%%%%%%%%%%%%%%%%%%%%
   \title{Parameter Effects in Circumplanetary Disk Spectra and Prospects for Spectral Fitting}

   % \subtitle{Subtitle}

%%%%%%%%%%%%%%%%%%%%%%%%%%%%%%%%%%%%%%%%
% Please separate each author with the \and command
%
% Please do not include ORCIDs next to author names.
% Only ORCIDs authenticated by individual authors in EDPS
% editorial system will be taken into account.
% ORCIDs included here will be removed.
%%%%%%%%%%%%%%%%%%%%%%%%%%%%%%%%%%%%%%%%

   \author{Xilei Sun\inst{\ref{SYSUSPhA}}\fnmsep\thanks{sunxlei@mail2.sysu.edu.cn}
        \and Gabriel-Dominique Marleau\inst{\ref{DUE},\ref{MPIA},\ref{BernWP}}
        \and Shang-Fei Liu\inst{\ref{SYSUSPhA},\ref{SYSUCSST}}\fnmsep\thanks{liushangfei@mail.sysu.edu.cn}
        }

   \institute{School of Physics and Astronomy, Sun Yat-sen University, 519082 Zhuhai, PR China%
   \label{SYSUSPhA}
   \and Fakult\"at f\"ur Physik, Universit\"at Duisburg--Essen, Lotharstra\ss{}e 1, 47057 Duisburg, Germany%
   \label{DUE}
   \and Max-Planck-Institut f\"ur Astronomie, K\"onigstuhl 17, 69117 Heidelberg, Germany%
   \label{MPIA}
   \and Division of Space Research \&\ Planetary Sciences, Physics Institute, University of Bern, Gesellschaftsstr.~6, 3012 Bern, Switzerland%
   \label{BernWP}
   \and CSST Science Center for the Guangdong-Hong Kong-Macau Greater Bay Area, Sun Yat-sen University, 519082 Zhuhai, PR China%
   \label{SYSUCSST}
   }

   % \date{Received September 30, 20XX}

% \abstract{}{}{}{}{}
% 5 {} token are mandatory
 
  \abstract
   {With the commissioning of the James Webb Space Telescope (JWST), near- and mid-infrared observations are rapidly extending into the wavelength regime where emission from small dust grains in circumplanetary disks (CPDs) is expected to dominate.}
   {We aim to systematically investigate how individual physical parameters of CPDs shape their infrared spectra and to improve the robustness of spectral fitting and physical interpretation of current and future observations.} 
   {Building on our previous parametric CPD models, we employ a parameter-grid approach combined with radiative transfer simulations to explore the dependence of observable spectra on disk structure and dust properties.}
   {We identify the physical mechanisms responsible for the main spectral features and parameter degeneracies, and present the global trends emerging from the parameter study. We also demonstrate the applicability of the models by fitting representative observational data.}
   {Our results provide a structured theoretical framework for interpreting near- and mid-infrared observations of CPDs with JWST and related facilities.}


   \maketitle

%%%%%%%%%%%%%%%%%%%%%%%%%%%%%%%%%%%%%%%%%%%%%%%%%%%%%%%%%%%%%%
\section{Introduction} 

    Planet formation naturally gives rise to compact structures of gas and dust surrounding giant planets, commonly referred to as circumplanetary disks (CPDs; \citealt{Ward2010,Tanigawa2012}). CPDs play a crucial role in the formation and evolution of planetary systems. They provide a continuous reservoir for mass accretion onto growing giant planets, thereby influencing the planets' final masses and thermal structures \citep{Ward2010, Zhu2015}. In addition, CPDs serve as the formation sites of planetary satellites, where their dynamical and thermodynamical environments regulate the emergence of Galilean-type moon systems \citep{Canup2002, Canup2006, Szulagyi2017, Fung2019, Batygin2025}. Understanding the structure, evolution, and radiative properties of CPDs is therefore essential for constructing a comprehensive picture of planet formation.

    With the advent of high-angular-resolution observational facilities, particularly ALMA and JWST, it has become increasingly feasible to search for forming planets and their surrounding CPDs \citep{Isella2019, Benisty2021}. Nevertheless, the observational identification of CPDs remains highly challenging \citep{Zhu2015}. CPDs are typically only a few astronomical units in size or smaller, making their emission easily overwhelmed by that of the surrounding protoplanetary disk (PPD). In addition, CPDs are generally cold and faint, imposing stringent requirements on both sensitivity and spatial resolution. Although several young systems, such as PDS~70 and AS~209, have exhibited structural or photometric features potentially associated with CPDs \citep{Benisty2021, Bae2022, Blakely2025}, the current evidence remains limited, and the physical properties of CPDs are still poorly constrained. An alternative observational window is provided by planetary-mass companions (PMCs), many of which are known to host circumplanetary-scale disks. Observations of systems such as GQ~Lup~B, TWA~27b, and YSES~1~b are relatively more accessible, and their surrounding disks can be regarded, in a broad sense, as CPDs \citep{Luhman2023, Cugno2024, Hoch2025}. These systems therefore offer valuable complementary insights into CPD formation and evolution.

    The formation, thermal structure, and radiative properties of CPDs have been extensively investigated from a theoretical perspective. One line of work relies on three-dimensional hydrodynamical simulations to study gap opening, accretion flows, and the dynamical structures arising from planet--disk interactions \citep{Dangelo2008, Tanigawa2012, Szulagyi2017, Fung2019, Schneeberger2024,Krapp2024,Schulik2025,Sagynbayeva2025}. Another line focuses on radiative transfer calculations, constructing parametric models with varying CPD masses, temperature structures, and dust compositions to predict their spectral and imaging signatures \citep{Isella2014, Zhu2015, Szulagyi2018,Portilla2022,Chen2022,Portilla2023,Taylor2025}. Our previous work primarily followed the latter approach, while incorporating constraints motivated by hydrodynamical simulations, with the goal of more directly connecting theoretical modeling to observational data \citep{Sun2024}.

    Despite this progress, parametric CPD models still face several key challenges \citep{Isella2014, Zhu2015, Szulagyi2018}. In particular, it remains unclear how individual model parameters shape observable CPD spectra and how these parameters should be adjusted to robustly match observational data, given the strong degeneracies among disk mass, temperature structure, and dust properties \citep{Isella2019, Benisty2021, Bae2022, Cugno2024}. Motivated by these limitations, this study extends our previous efforts by introducing a more physically consistent atmospheric structure into the parametric CPD framework and by employing a parameter-grid approach to systematically quantify the influence of different parameters on the resulting spectral features. This framework is designed to provide more robust theoretical guidance for future CPD detection and interpretation. Drawing inspiration from atmospheric modeling studies of brown dwarfs and planetary-mass objects \citep{Burrows2006, Allard2012, Spiegel2012, Marley2015}, we expect that our improved model will yield more accurate predictions for CPD observations and serve as a more powerful tool for fitting observed spectra.

    This paper is organized as follows. In Section~\ref{sec:methods}, we describe the parameter-grid framework adopted for the CPD models and the details of the radiative transfer simulations. Section~\ref{sec:results} presents the main results of the parameter exploration. Section~\ref{sec:applications} demonstrates applications of the models to observational data. In Section~\ref{sec:discussion}, we discuss the physical implications of our results and the underlying mechanisms of the models. Finally, Section~\ref{sec:conclusions} summarizes our main findings and conclusions.

%%%%%%%%%%%%%%%%%%%%%%%%%%%%%%%%%%%%%%%%%%%%%%%%%%%%%%%%%%%%%%
\section{Methods} \label{sec:methods}

    Radiative transfer modelling of unresolved disks has been extensively applied in the context of circumstellar disks around T Tauri stars, very low-mass stars, and brown dwarfs (e.g., \citealp{Dullemond2001, Whitney2003a, Whitney2023b, Whitney2004, Pinte2006, Espaillat2014}). These studies have demonstrated the importance of disk geometry, dust properties, and irradiation in shaping the observed spectral energy distributions. Building upon this general framework, we focus here on circumplanetary disks and systematically investigate how variations in key physical parameters affect their infrared spectra.

    To this end, we construct a series of parametric CPD models spanning a range of physically motivated parameter values. The model grid is designed to enable a systematic exploration of parameter dependencies, with particular emphasis on spectral signatures in the near- and mid-infrared, motivated by the growing availability of observations at these wavelengths. We then perform radiative transfer simulations to compute the corresponding synthetic spectra.

    \begin{figure*}[t!]
        \centering
        \includegraphics[width=0.7\textwidth]{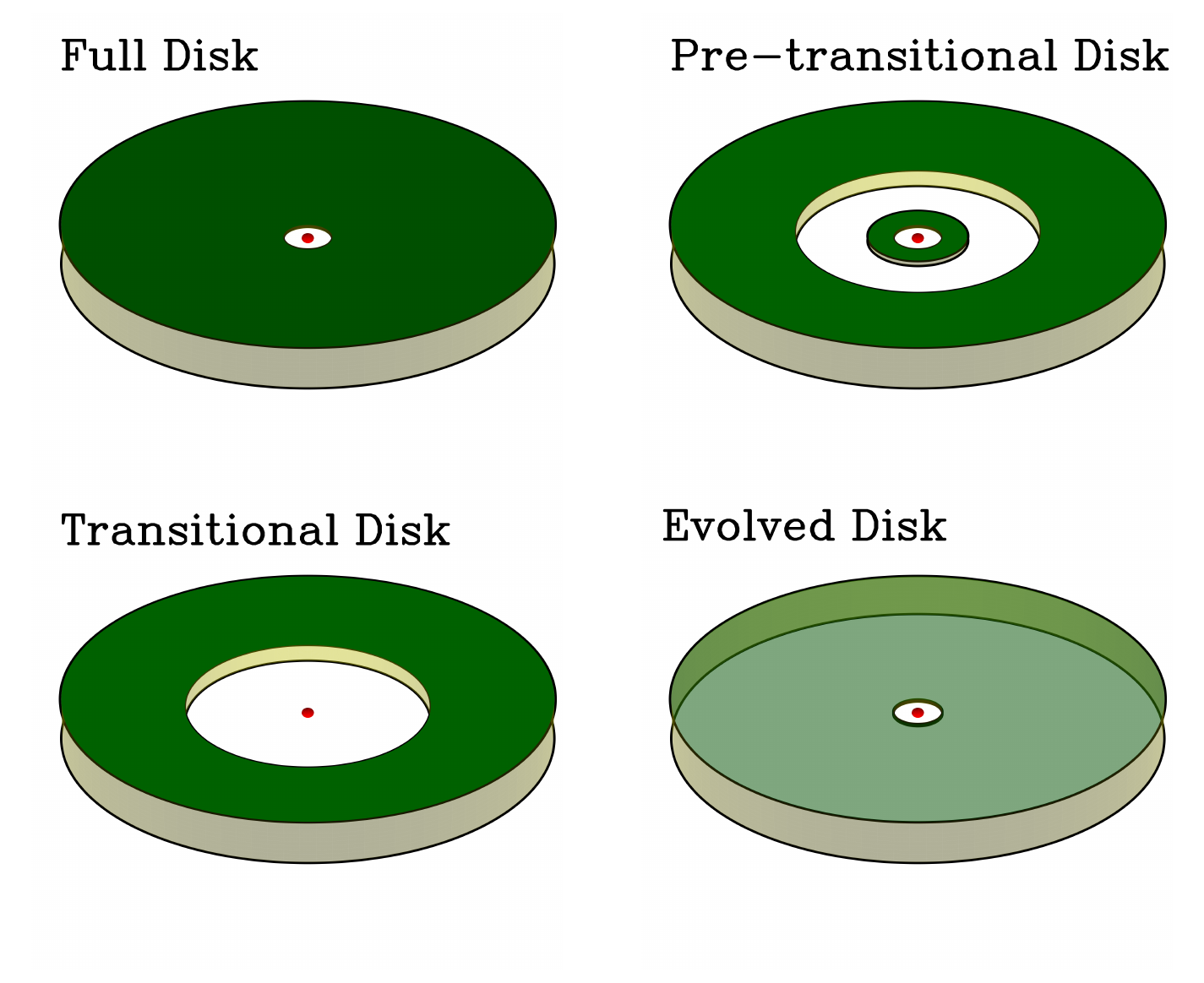}
        \caption{Schematic diagrams of the four disk types are shown. These are illustrated with an inclination for clarity, but all spectral results assume $i=0^\circ$.}
        \label{fig:1}
    \end{figure*}

    \subsection{CPD model setups} \label{subsec:tables}

        Following our previous work \citep{Sun2024}, we assume that each CPD model consists of a central planet surrounded by a dusty circumplanetary disk. For simplicity, we neglect the contribution from the gas component, as dust emission is expected to dominate the continuum radiation in the infrared wavelength range considered here. We adopt a representative planetary mass of $M_\mathrm{P} = 10\,M_\mathrm{J}$, a planetary radius of $R_\mathrm{P} = 2\,R_\mathrm{J}$, and an effective temperature of $T_\mathrm{P} = 2000\,\mathrm{K}$, where $M_\mathrm{J}$ and $R_\mathrm{J}$ denote the mass and radius of Jupiter, respectively.
        
        For the full disk prototype, the radial dust surface density profile is described by
        \begin{equation}\label{eq:1}
            \Sigma_\mathrm{d} (R) = \Sigma_\mathrm{d,0} \left( \frac{R}{R_0} \right) ^ \gamma,
        \end{equation}
        where $R$ represents the radius, $\Sigma_\mathrm{d,0}$ represents the dust surface density at $R=R_0$, and $\gamma$ is the density power-law index. Simultaneously, in the vertical direction, the density distribution is assumed to follow the Gaussian vertical profile
        \begin{equation}\label{eq:2}
            \rho_\mathrm{d} (R, z) = \frac{\Sigma_\mathrm{d}(R)}{\sqrt{2\pi}\, h_\mathrm{d}(R)}\:\mathrm{exp}\left[-\frac{1}{2} \left(\frac{z}{h_\mathrm{d}(R)}\right)^2\right],
        \end{equation}
        where $R$ and $z$ represent the radius and height in cylindrical coordinates, respectively. $h_\mathrm{d}$ represents the dust scale height. The radial dependence of the dust scale height is parameterized as
        \begin{equation}\label{eq:3}
            h_\mathrm{d}(R)=h_\mathrm{d,0} \left( \frac{R}{R_0} \right) ^ \beta,
        \end{equation}
        where $h_\mathrm{d,0}$ is the dust scale height at distance $R_0$, and $\beta$ is the flaring index.

        We adopt the same classification scheme for CPD models as in our previous work \citep{Sun2024}, which are \textit{full disk}, \textit{pre-transitional disk}, \textit{transitional disk} and \textit{evolved disk} (schematic diagrams, see Fig.~\ref{fig:1}). For completeness, we summarize below both the conceptual descriptions and the corresponding dust surface density prescriptions for the four disk prototypes:
        \begin{itemize}
            \item 
            A \textbf{full disk} has a continuous and smooth dust distribution without large-scale substructures.
            \item 
            A \textbf{pre-transitional disk} is similar to a full disk but contains a gap, i.e., a radially localized region where the dust density is significantly depleted.
            \item 
            A \textbf{transitional disk} features a central cavity, such that the inner disk is largely cleared and the disk inner edge is located at a larger distance from the planet.
            \item 
            A \textbf{evolved disk} can be understood as a full disk with a uniformly reduced (scaled-down) dust surface density, representing a more advanced evolutionary stage.
        \end{itemize}
        \begin{align}
            \Sigma_\mathrm{d,F} (R) &= \Sigma_\mathrm{d,0}  \left( \frac{R}{R_0} \right) ^ \gamma,\\
            \Sigma_\mathrm{d,P} (R) &= \begin{cases}\Sigma_\mathrm{d,F} (R) \cdot \left( \frac{\Sigma_\mathrm{gap}}{\Sigma_\mathrm{d}} \right),\quad & R_{\mathrm{in}} \le R \le R_{\mathrm{out}}\\ \Sigma_\mathrm{d,F} (R), \quad & R<R_{\mathrm{in}}\:\&\: R > R_{\mathrm{out}}
            \end{cases},\\
            \Sigma_\mathrm{d,T} (R) &= \begin{cases}\Sigma_\mathrm{d,F} (R) \cdot \left( \frac{\Sigma_\mathrm{cavity}}{\Sigma_\mathrm{d}} \right),\quad & R \le R_{\mathrm{cavity}}\\ \Sigma_\mathrm{d,F} (R), \quad & R > R_{\mathrm{cavity}} 
            \end{cases},\\
            \Sigma_\mathrm{d,E} (R) &= \Sigma_\mathrm{d,F} (R) \cdot \left( \frac{\Sigma_\mathrm{E}}{\Sigma_\mathrm{d}} \right),
        \end{align}
        where the subscripts F, P, T and E indicate each prototype, and $\Sigma_\mathrm{gap}$, $\Sigma_\mathrm{cavity}$, and $\Sigma_\mathrm{E}$ denote scaling factors that control the dust depletion level in each prototype.
        
        We select a set of representative values for each model parameter to construct a parameter grid, as summarized in Table~\ref{tab:model}.

        % \onecolumn
        \begin{table*}
            \caption{Parameters of Circumplanetary Disk Models}
            \centering
            \begin{tabular}{lcr}
                \hline\hline
                \textbf{Parameter} & \textbf{Symbol}& \textbf{Value} \\
                \hline
                \textbf{Common Parameters\tablefootmark{a}} & \\
                \hline
                \textbf{Planet (BT-Settl CIFIST)} \\
                Effective temperature&$T_{\rm eff}$ & $2000\:\mathrm{K}$\\
                Mass & $M_\mathrm{p}$ & $10\: M_\mathrm{J}$\\
                Radius & $R_\mathrm{p}$ & $2\: R_\mathrm{J}$\\
                \\
                \textbf{Dust grains (DSHARP opacity)}\\
                Grain size range\tablefootmark{b} & $[a_\mathrm{min},a_\mathrm{max}]$ & $[0.1,\:10]\:\upmu\mathrm{m}$\\
                Grain size power index & $q$ & 3.5 \\
                \\
                \textbf{Dust disk}\\
                Reference radius &$R_0$ & $10\:R_\mathrm{J}$\\
                Inner radius &$R_\mathrm{in}$ & $5\:R_\mathrm{J}$\\
                Outer radius &$R_\mathrm{out}$ & $50,\:275,\:\mathbf{500},\:725,\:950\:R_\mathrm{J}$\\
                Dust surface density at $R_0$ & $\Sigma_\mathrm{d,0}$ & $0.1,\:0.3,\:\mathbf{0.5},\:0.7,\:0.9 \: \mathrm{g \cdot cm^{-2}}$ \\
                Density power-law index&$\gamma$ & $-1.4,\:-1.2,\:\mathbf{-1.0},\:-0.8,\:-0.6$\\
                Aspect ratio at $R_0$ & $h_\mathrm{d,0}/R_0$ & $0.03,\:0.05,\:\mathbf{0.07},\:0.09,\:0.11$\\
                Flaring index &$\beta$ & $1.1,\:1.15,\:\mathbf{1.2},\:1.25,\:1.3$\\
                \\
                \textbf{Mesh}\\
                Radial grid range& $R$ & $\left[R_\mathrm{in},\:R_\mathrm{out}\right]$\\
                Azimuthal grid range& $\phi$ & $\left[0,\:2\pi\right)$\\
                Polar grid range&$\theta$ & $\left[0,\:\frac{\pi}{6}\right]$\\
                Mesh resolution & $N_R \times N_\phi \times N_\theta $ &$100\:\times\:1\:\times\:32$\\
                \\
                \hline
                \textbf{Type-specific parameters\tablefootmark{c}} &  \\
                \hline
                \textbf{Pre-transitional disk} \\
                Gap inner edge &$R_\mathrm{gap,in}$ & $\mathbf{10},\:30,\:50\:R_\mathrm{J}$\\
                Gap outer edge &$R_\mathrm{gap,out}$ & $20,\:40,\:\mathbf{60},\:80,\:100\:R_\mathrm{J}$\\
                Gap radial range &$R_\mathrm{gap}$ & $\left[R_\mathrm{gap,in},\:R_\mathrm{gap,out}\right]$\\
                Gap depletion factor & $\Sigma_{\mathrm{gap}}/\Sigma_\mathrm{d}$ & $10^{-4},5\times10^{-4},\mathbf{10^{-3}},5\times10^{-3},10^{-2}$\\
                \\
                \textbf{Transitional disk}\\
                Cavity outer edge &$R_\mathrm{cavity}$ & $25,\:50,\:\mathbf{100},\:200,\:400\:R_\mathrm{J}$\\
                Cavity depletion factor & $\Sigma_{\mathrm{cavity}}/\Sigma_\mathrm{d}$ & $10^{-4},5\times10^{-4},\mathbf{10^{-3}},5\times10^{-3},10^{-2}$\\
                \\
                \textbf{Evolved disk}\\
                Overall depletion factor& $\Sigma_\mathrm{E}/\Sigma_\mathrm{d}$ & $10^{-4},5\times10^{-4},\mathbf{10^{-3}},5\times10^{-3},10^{-2}$\\
                \hline\hline
            \end{tabular}\\
            \label{tab:model}
            \tablefoot{Values in bold are used in the fiducial model.\\
                \tablefoottext{a}{Parameters in this section are applied to all four disk types.}\\
                \tablefoottext{b}{This study focuses on near-to-mid-infrared wavelengths, which are mostly sensitive to small grains up to 10 $\upmu$m.}\\
                %%\tablefoottext{c}{The bold value is used in the fiducial model, as below.}\\
                \tablefoottext{c}{Parameters in this section are applied to specific disk types.}}
        \end{table*}

    % \twocolumn
    \subsection{Fiducial model parameters}

        To facilitate a systematic investigation of the influence of individual parameters, we define a set of fiducial models that serve as reference points within the parameter grid. Because the four disk prototypes considered in this work exhibit distinct structural characteristics, we construct one fiducial model for each prototype, resulting in a total of four fiducial models. The adopted fiducial parameter values are highlighted in boldface in Table~\ref{tab:model}.

        At present, both observational constraints and theoretical predictions for CPDs remain highly uncertain, leaving the physically plausible parameter space relatively broad. In this context, we deliberately choose the fiducial parameter values to lie near the central region of the explored parameter space, rather than at its extremes. This choice is intended to provide representative reference models while avoiding bias toward extreme or highly specific configurations. The primary goal of this study is not to identify a unique best-fitting model, but to quantify how variations in individual parameters affect the resulting CPD spectra. For this purpose, the adopted fiducial models and parameter grid are sufficient to capture the main parameter dependencies relevant for current observations.

    \subsection{Radiative Transfer Simulations}

        We perform radiative transfer simulations for all CPD models using \texttt{RADMC-3D} \citep{Dullemond2012}. In contrast to our previous work, where the planetary radiation was approximated as a blackbody, we adopt realistic atmospheric models for planetary-mass objects generated with \texttt{species} \citep{Stolker2020}. This approach provides more physically consistent input spectra and enhances the reliability of the simulated disk emission. In this study, we adopt the \texttt{BT-Settl-CIFIST} model \citep{Allard2012} as a representative atmospheric template, noting that alternative atmospheric models produce only minor variations in the MIR-band resulting spectra, where the CPD emission dominates.

        The dust properties follow the same setup as in our previous work. In this work, we consider only small dust grains that are well coupled to the gas and dominate the opacity at near- and mid-infrared wavelengths. Larger grains, such as millimeter-sized pebbles that are expected to settle toward the disk midplane, are not included in our models. While such grains can contribute significantly to the emission at (sub-)millimeter and radio wavelengths, their inclusion would require a more complex treatment of grain growth, vertical settling, and potentially optical depth effects. As a result, the present models are primarily intended to describe the infrared spectral properties of circumplanetary disks, and their predictions are not expected to be directly applicable in the radio regime. We adopt a grain size range of $a \in [0.1,\,10]\,\upmu\mathrm{m}$ with a power-law distribution $n(a) \propto a^{-q}$ and $q = 3.5$, representing typical small grains in CPDs. The dust composition is based on the DSHARP mixture \citep{Birnstiel2018}, including silicates and carbonaceous materials, and the absorption and scattering opacities are computed using the \texttt{dsharp\_opac} routine within the Mie theory framework.

        The computational grid is defined in spherical coordinates with dimensions $100 \times 1 \times 32$ in the radial, azimuthal, and polar directions, respectively. The azimuthal direction contains only one grid cell, justified by the near-axisymmetric structure of CPDs in this study. The radial grid extends from $R_{\mathrm{in}}$ to $R_{\mathrm{out}}$ (see Table~\ref{tab:model}), while the polar grid covers $-30^\circ$ to $30^\circ$, encompassing the main vertical extent of the dust disk. Symmetry with respect to the radial–azimuthal plane is assumed to further reduce computational cost.

        We employ $5 \times 10^5$ photon packets for ray tracing and compute the temperature structure using the thermal Monte Carlo method. Spectral outputs are subsequently generated based on the resulting temperature distributions.The spectra are scaled assuming a source distance of 50 pc, which sets the absolute flux level for comparison with observational data. To isolate clearly the effects of individual parameters in our grid study, scattering is neglected in these initial spectral calculations, and all models are assumed to be observed face-on. The influence of scattering and different viewing inclinations is examined separately in Section~\ref{sec:scattering}.

%%%%%%%%%%%%%%%%%%%%%%%%%%%%%%%%%%%%%%%%%%%%%%%%%%%%%%%%%%%%%%
    \begin{figure*}[t!]
        \centering
        \includegraphics[width=0.9\textwidth]{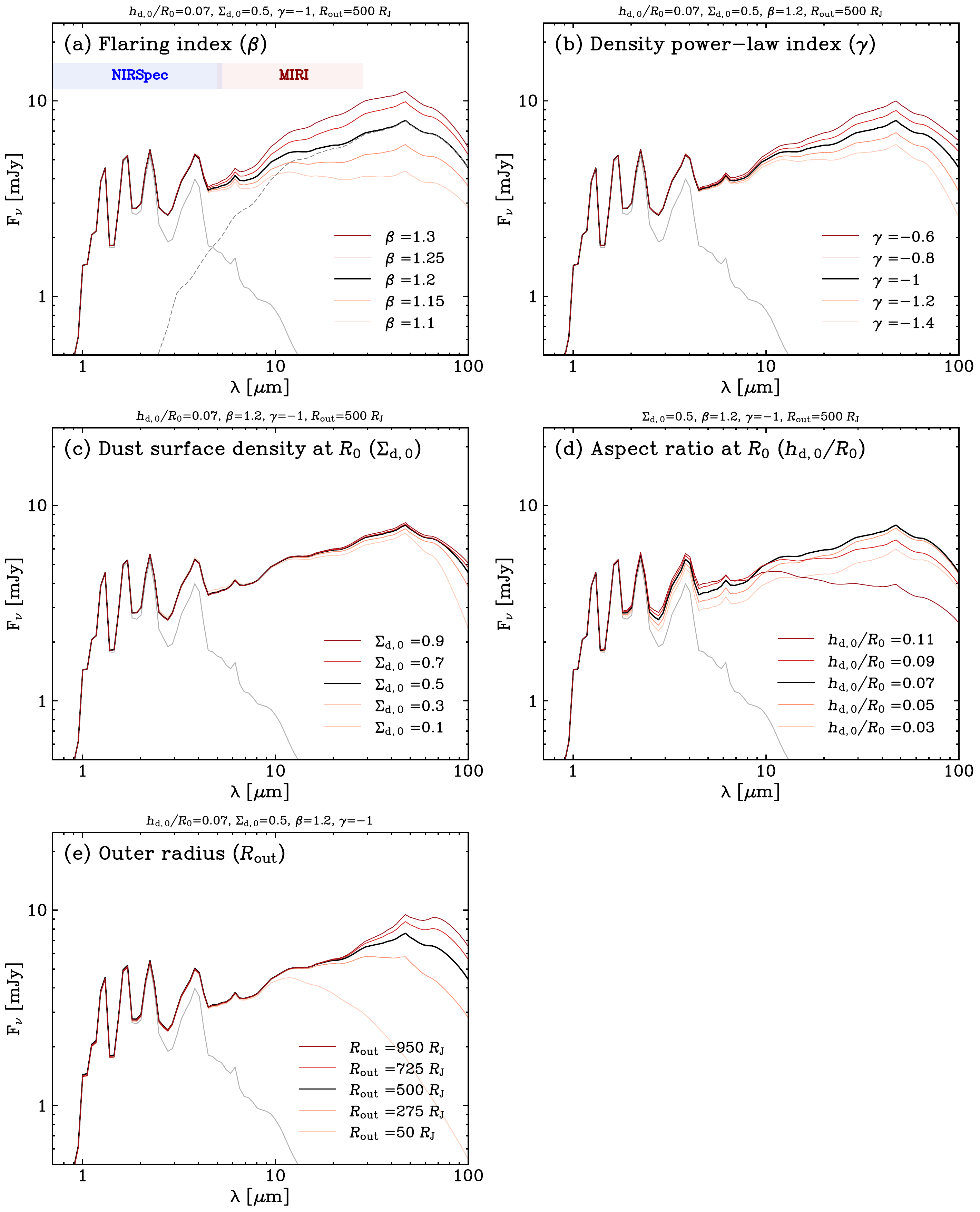}
        \caption{Spectral results from the parameter grid for the full disk models. Colored lines show models with different values of the varied parameter in each panel, while the black line denotes the fiducial model. The gray line shows the contribution from the planetary atmosphere alone. All panels share the fiducial parameters listed in Table~\ref{tab:model}, except for the parameter varied in each panel. In panel~a, the dashed line shows the disk contribution in the fiducial model, and the blue and red shaded regions indicate the JWST/NIRSpec and MIRI bands, respectively.}
        \label{fig:2}
    \end{figure*}

\section{Results} \label{sec:results}

    In this section, we present the spectral results obtained from the parameter-grid exploration, highlighting their dependence on individual model parameters through comparisons with the fiducial models. Results for different CPD prototypes are shown separately, while the physical interpretation of these trends is deferred to Section~\ref{sec:discussion}.

    \begin{figure*}[t!]
        \centering
        \includegraphics[width=0.9\textwidth]{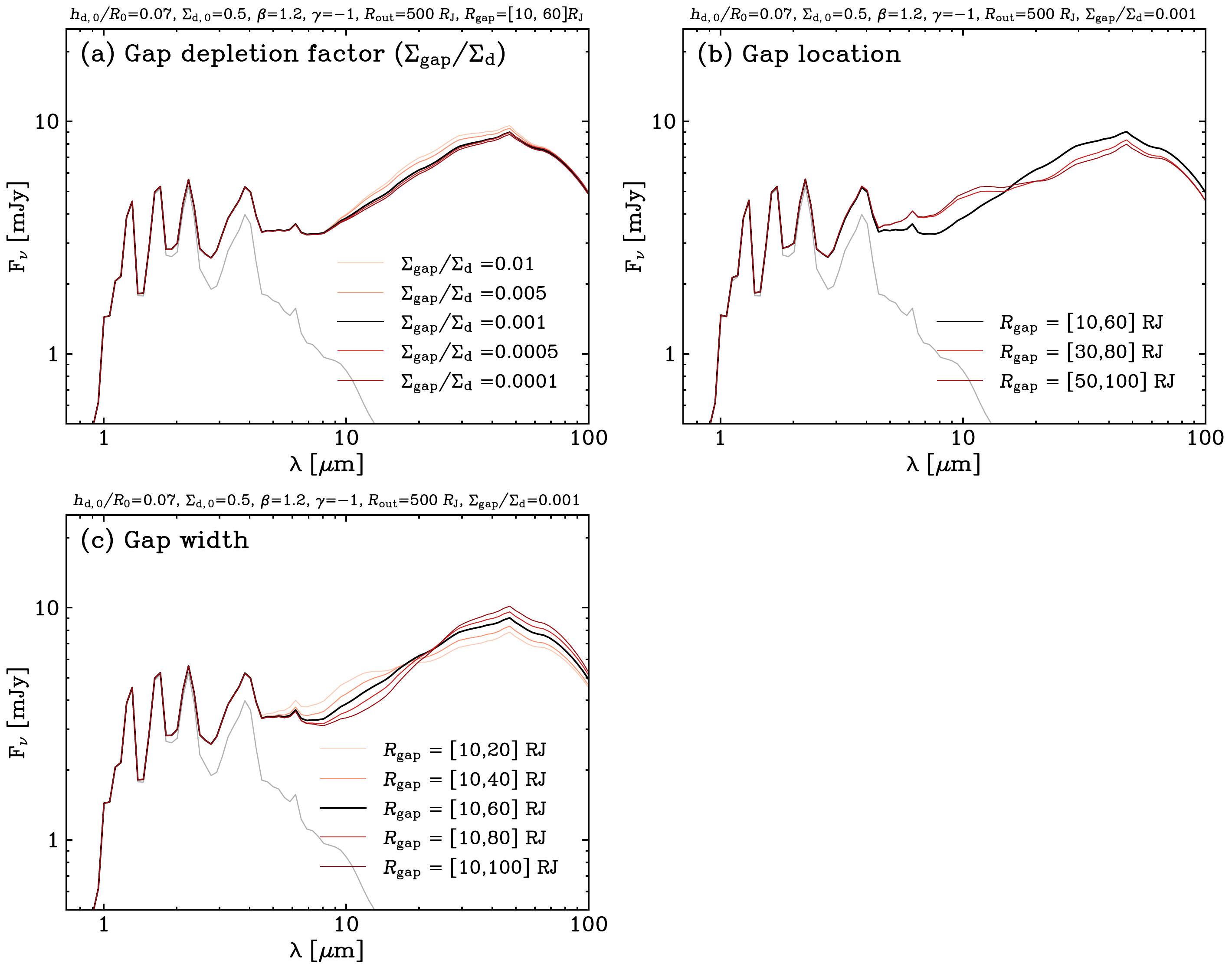}
        \caption{Same as Figure~2, but showing the spectral results from the parameter grid method for pre-transitional disks.}
        \label{fig:3}
    \end{figure*}

    \subsection{Full disks}

        We first consider the simplest configuration, the full disk CPD model. The influence of magnetospheric accretion on the inner boundary $R_{\mathrm{in}}$ is deferred to the discussion of transitional disks. For the full disk case, we construct a parameter grid spanning five key parameters that primarily control the total dust mass and its spatial distribution: flaring index ($\beta$), aspect ratio at $R_0$ ($h_\mathrm{d,0}/R_0$), density power-law index ($\gamma$), Dust surface density at $R_0$ ($\Sigma_{\mathrm{d},0}$), and the outer radius ($R_{\mathrm{out}}$). Although the full parameter space has been explored, for clarity we present here only the subset of models in the vicinity of the fiducial model, as shown in Figure~\ref{fig:2}. The black line denotes the fiducial model, colored lines represent models with different parameter values, and the gray line shows the contribution from the planetary atmosphere. In Figure~\ref{fig:2}a, the dashed line shows the contribution from the disk in the fiducial model.

        In the infrared regime, the full disk models are optically thick at short wavelengths, such that the emergent thermal emission is dominated by dust near the disk surface. At longer wavelengths, the dust opacity decreases, and the emission originates from progressively deeper layers. In addition, the peak wavelength of the dust thermal emission depends on temperature, making the resulting spectral variations a combined effect of both optical depth and temperature structure.

        The flaring index $\beta$ controls the overall curvature of the disk. A larger $\beta$ produces a more flared geometry, enabling the disk surface to intercept a greater fraction of the planetary radiation and thereby increasing the dust temperature. This effect is most pronounced in the mid- to outer disk regions, where the dust is relatively cold. As a consequence, the spectral flux increases at wavelengths longer than $\sim 5\,\upmu\mathrm{m}$, as illustrated in Fig.~\ref{fig:2}a.

        The density power-law index $\gamma$ determines how dust mass is radially distributed between the inner and outer disk. A steeper slope shifts more mass toward larger radii. In our models, the reference radius is fixed at $R_0 = 10\,R_{\mathrm{J}}$. Because the inner disk is already optically thick, variations in $\gamma$ have little effect on the thermally emitting dust there. In contrast, the outer disk can be optically thin, such that an increased dust mass enhances the emission from colder material. As a result, changes in $\gamma$ mainly affect longer wavelengths and with a smaller amplitude than variations in the flaring index, as shown in Fig.~\ref{fig:2}b.

        The parameter $\Sigma_{\mathrm{d},0}$ in Equation~\ref{eq:1} sets the normalization of the dust surface density and therefore the total dust mass, while preserving the relative spatial distribution. For the optically thick full disk models, varying $\Sigma_{\mathrm{d},0}$ has only a modest impact on the emergent spectrum, primarily at long wavelengths. Even when the total dust mass differs by an order of magnitude, the resulting spectral differences remain small, as shown in Fig.~\ref{fig:2}c.

        The aspect ratio $h_{\mathrm{d},0}/R_0$ controls the geometrical thickness of the disk. While it does not directly modify the face-on optical depth for a fixed surface density, it significantly alters the vertical density structure and temperature distribution near the disk surface. Increasing the aspect ratio leads to a thicker disk, allowing the inner regions to intercept more radiation and become warmer, while the outer disk is increasingly shielded and cools. This results in enhanced short-wavelength emission and reduced long-wavelength flux. In contrast, decreasing the aspect ratio lowers the overall disk temperature, leading to a reduction in flux over most wavelengths. The resulting spectral response is therefore non-linear, reflecting the coupled effects of disk geometry and radiative heating, as shown in Fig.~\ref{fig:2}d.

        The outer radius $R_{\mathrm{out}}$ sets the radial extent of the disk and has the most straightforward effect on the spectrum. For fixed surface density and scale-height profiles, a larger $R_{\mathrm{out}}$ corresponds to a more extended disk containing a greater mass of cold dust, thereby enhancing the flux at long wavelengths. The wavelength range and magnitude of this enhancement depend on the specific value of $R_{\mathrm{out}}$, as shown in Fig.~\ref{fig:2}e.

        \begin{figure*}[t!]
            \centering
            \includegraphics[width=0.9\textwidth]{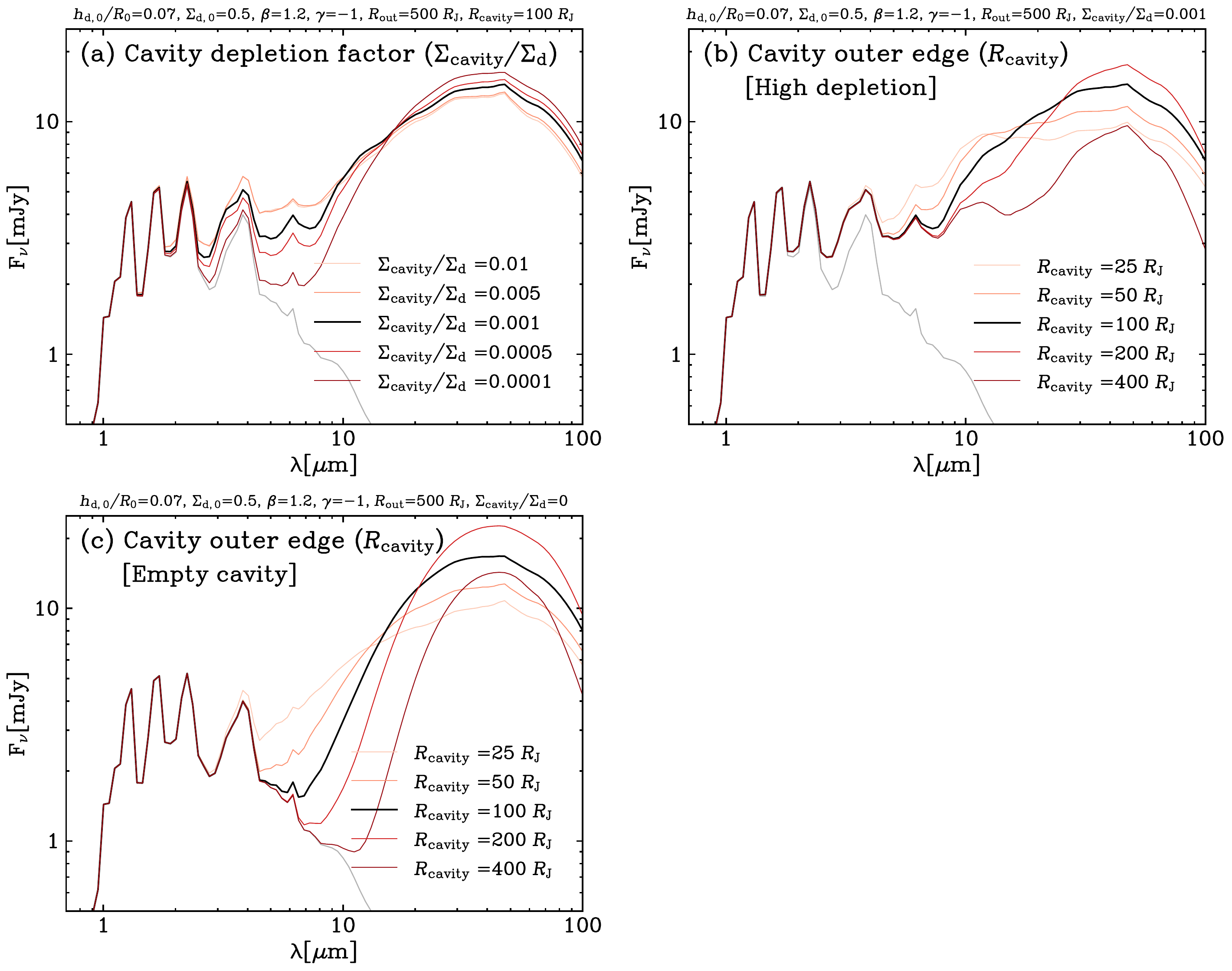}
            \caption{Same as Figure~2, but showing the spectral results from the parameter grid method for transitional disks. Panels (b) and (c) show models with different cavity depletion factors ($10^{-3}$ and $0$, respectively).}
            \label{fig:4}
        \end{figure*}

    \subsection{Pre-transitional Disks}

        In this section, we examine the intermediate case of pre-transitional disks. These models are characterized by the presence of a gap separating an inner and an outer dust disk. In our setup, we consider only a single gap. Although real disks may host multiple gaps and rings, the spectral impact of such complex structures can, to first order, be approximated as the superposition of several single-gap configurations. We therefore focus on the single-gap case for clarity.

        The key parameters governing pre-transitional disks are associated with the gap geometry, namely its depth, radial extent, and location. These properties are controlled by the \textit{gap depletion factor}, the \textit{gap inner edge}, and the \textit{gap outer edge}, respectively.

        We first examine the effect of the gap depletion factor, which sets the dust density within the gap and thus its optical depth. A lower depletion factor reduces the amount of dust available to emit thermal radiation in the gap region, leading to a decrease in the spectral flux at wavelengths primarily originating from that region. This behavior is illustrated in Fig.~\ref{fig:3}a and is consistent with physical expectations.

        Next, we investigate the influence of the gap location by varying the gap inner edge while keeping the gap width fixed. Because different radial regions of the disk correspond to different characteristic temperatures, shifting the gap outward affects the spectrum in a wavelength-dependent manner. When the gap is located closer to the planet, it suppresses emission at shorter wavelengths. As the gap moves to larger radii, the temperature of the affected region decreases, and the resulting spectral deficit shifts toward longer wavelengths. This trend is shown in Fig.~\ref{fig:3}b.

        Finally, we explore the effect of the gap width by fixing the gap inner edge at $10\,R_{\mathrm{J}}$ and progressively increasing the gap width. A wider gap produces a broader low-density and optically thin region, resulting in a more extended spectral dip. However, the response is not purely monotonic. Even in the presence of a gap, a \textit{warm wall} forms at the outer edge of the gap due to direct irradiation. Because pre-transitional disks retain an inner dust disk, the temperature of this warm wall is significantly lower than that in transitional disks, yet it still contributes a modest excess at longer wavelengths. As the gap width increases, the relative contribution of this warm wall becomes slightly more pronounced, leading to the behavior seen in Fig.~\ref{fig:3}c.

    \subsection{Transitional Disks}

        We now turn to a distinct class of models: transitional disks. These disks are characterized by an inner cavity, which may arise from a variety of physical mechanisms, including magnetospheric accretion, photoevaporation, or satellite formation. Owing to the presence of the cavity, transitional disks often exhibit a pronounced rise in their mid-infrared spectra, commonly attributed to thermal emission from the warm inner wall at the cavity outer edge. In this class of models, the dominant parameters are the \textit{cavity depletion factor} and the \textit{cavity outer edge}, which respectively control the depth and radial extent of the cavity.

        In contrast to evolved disks, the depletion factor in transitional disks only modifies the dust density within the cavity region. Because the cavity is located in the inner disk and is typically optically thin, this parameter has a strong influence on the amount of dust capable of emitting thermal radiation at short wavelengths, and thus primarily affects the near-infrared portion of the spectrum. Once the inner cavity becomes optically thin, the cavity outer edge is more directly irradiated by the central source, giving rise to an additional \textit{warm wall}. Although this wall does not reach the temperatures attained when the inner disk is filled with dust, it nevertheless produces a noticeable excess in the mid-infrared. As the depletion factor decreases, the near-infrared flux progressively diminishes and approaches the level set by the planetary atmosphere, while the contribution from the warm wall becomes increasingly prominent in the mid-infrared, as illustrated in Fig.~\ref{fig:4}a.

        To further examine the role of the \textit{cavity outer edge}, we compare two scenarios shown in Fig.~\ref{fig:4}b and~c, which correspond to different assumptions for the surface density inside the cavity. Specifically, panel~b adopts a depletion factor of $10^{-3}$, representative of a typical transitional disk, while panel~c assumes a fully cleared cavity with a depletion factor of $0$. Although the same parameter, $R_{\mathrm{cavity}}$, is varied in both panels, the resulting spectral behavior differs due to these distinct assumptions. In general, the cavity outer edge determines the distance between the warm wall and the planet, thereby setting the wall temperature and the characteristic wavelength of its thermal emission. For small cavity radii, the warm wall remains relatively hot, and the resulting spectrum more closely resembles that of a full disk. As the cavity outer edge increases, the wall temperature decreases, shifting the peak of the warm-wall emission toward longer wavelengths and producing a more distinct transitional disk spectral signature.

        However, when the cavity outer edge becomes sufficiently large (e.g., $400\,R_{\mathrm{J}}$), the total dust mass—particularly in the vicinity of the warm wall—becomes too low to contribute substantial thermal emission. As a result, despite the continued shift of the warm-wall emission peak to longer wavelengths, the overall spectral flux decreases, as shown in Fig.~\ref{fig:4}b and~c.

    \subsection{Evolved Disks}

        In this section, we consider the evolved disk models, which represent a relatively simple extension of the full disk case. These models share the same basic parameter set as the full disk models, but correspond to highly evolved, nearly dissipated disks with substantially reduced dust densities throughout the disk. This reduction is characterized by a global \textit{depletion factor}, which serves as the only additional parameter introduced in this class of models. The effects of the remaining five parameters are qualitatively similar to those in the full disk models and are therefore not repeated here.

        \begin{figure}[t!]
            \centering
            \includegraphics[width=0.45\textwidth]{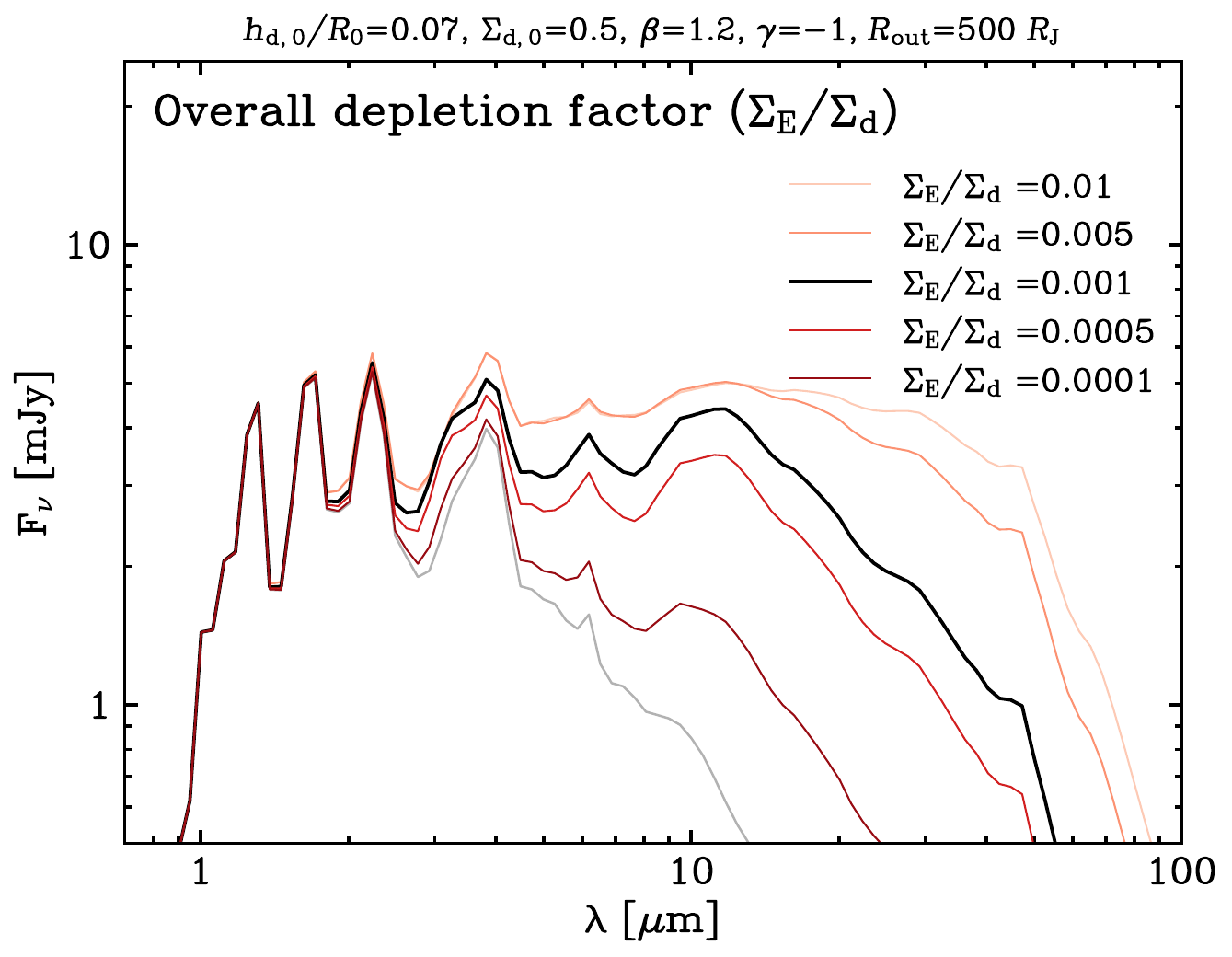}
            \caption{Same as Figure~2, but showing the spectral results from the parameter grid method evolved disks.}
            \label{fig:5}
        \end{figure}

        Because the depletion factor uniformly scales the dust surface density at all radii, its effect is in some sense analogous to that of $\Sigma_{\mathrm{d},0}$. However, unlike the optically thick full disk case, evolved disks are generally optically thin, such that their thermal emission is more directly proportional to the amount of dust available to absorb and reprocess the planetary radiation. Consequently, increasing the depletion factor leads to a systematic rise in spectral flux over the infrared wavelength range, as shown in Fig.~\ref{fig:5}. When the depletion factor becomes relatively large (e.g., $\gtrsim 0.01$), the resulting spectra increasingly resemble those of full disk models, reflecting the continuous transition between these disk classes.

%%%%%%%%%%%%%%%%%%%%%%%%%%%%%%%%%%%%%%%%%%%%%%%%%%%%%%%%%%%%%%
\section{Applications} \label{sec:applications}

    \begin{figure*}[t!]
        \centering
        \includegraphics[width=0.85\textwidth]{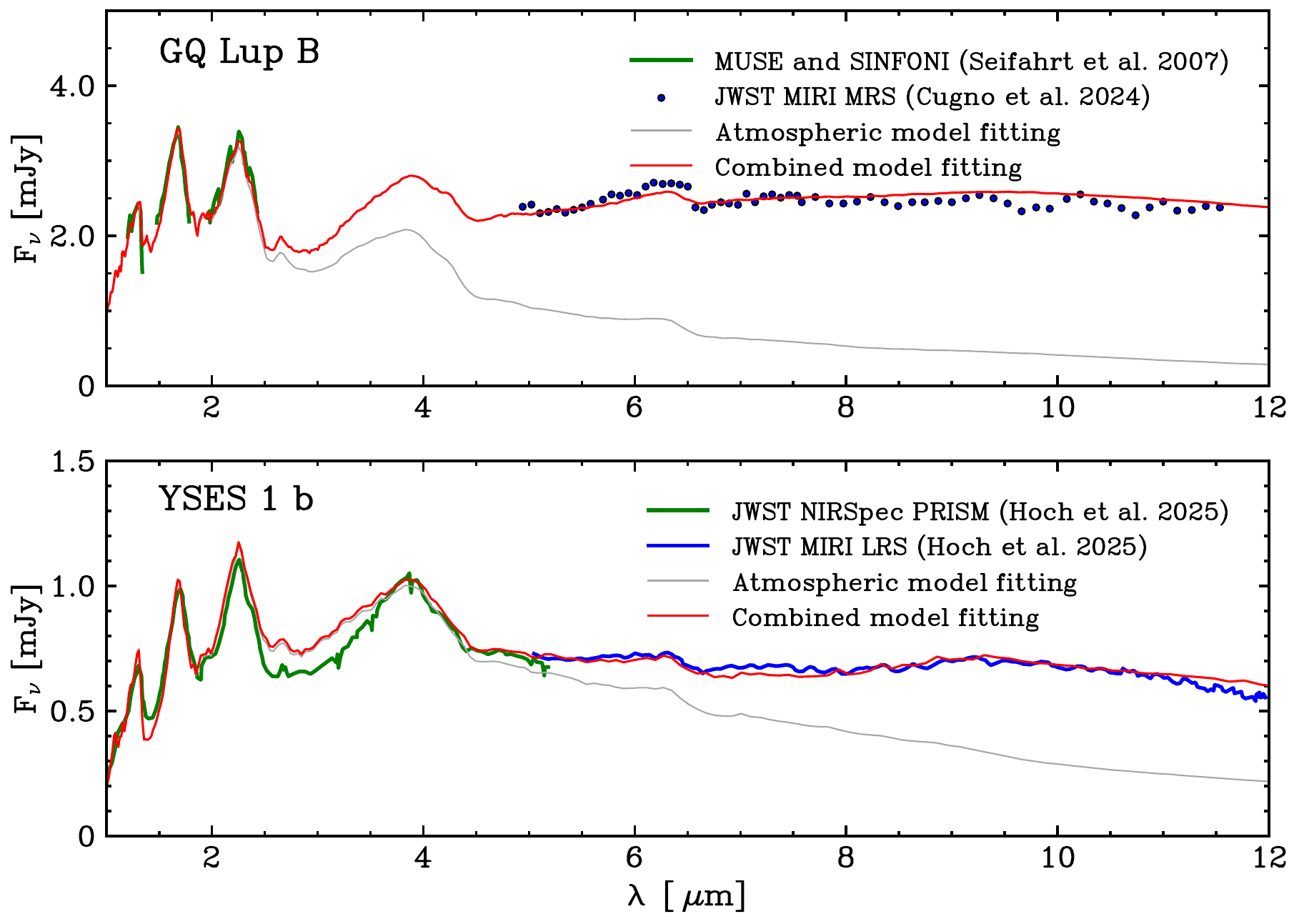}
        \caption{Spectral fitting results for GQ~Lup~B and YSES~1~b. \textbf{Top panel:} The green curve shows the near-infrared data from MUSE and SINFONI \citep{Seifahrt2007}, while the blue points represent the mid-infrared JWST/MIRI LRS data \citep{Cugno2024}. The gray curve denotes the atmospheric model used in the fitting, and the red curve shows the combined spectrum of the atmospheric and disk models. \textbf{Bottom panel:} The green and blue curves represent the near-infrared and mid-infrared data from JWST/NIRSpec and MIRI, respectively \citep{Hoch2025}. The gray and red curves correspond to the atmospheric model and the combined spectrum, respectively.}
        \label{fig:6}
    \end{figure*}

    Following the systematic exploration of parameter influences presented in the previous sections, we now demonstrate practical applications of our modeling framework by fitting several representative observational datasets. The goal of this section is not to derive definitive parameter constraints, but rather to illustrate how the parameter grid method can be used to interpret infrared excesses and to assess the plausibility of circumplanetary disk scenarios for individual sources.

    With the rapid advancement of infrared observational facilities, particularly JWST, an increasing number of spectra for planetary and planetary-mass companions have become available. In addition to molecular absorption and emission features that probe atmospheric composition, many of these objects exhibit infrared excesses at mid-infrared wavelengths. Such excesses provide valuable diagnostics of circumplanetary material and may encode information about disk geometry, evolutionary stage, and dust content.

    TWA~27~b is an L6 planetary-mass companion orbiting the M9 brown dwarf TWA~27~A. Recently, \citet{Patapis2025} presented reduced JWST/MIRI-MRS observations of this system. Applying our transitional disk model, they find that the spectral structure between 8~and 10\,$\upmu$m, as well as the photometric point near 15\,$\upmu$m, can be reproduced reasonably well. The resulting fit is consistent with the presence of a partially depleted inner region and a warm disk wall, a configuration naturally captured by the transitional disk framework (see Fig.~7 in \citealt{Patapis2025}). Possible silicate absorption features are also suggested by the data, although a detailed mineralogical analysis is beyond the scope of this work.

    GQ~Lup~B is an accreting brown dwarf companion for which \citet{Cugno2024} reported JWST/MIRI-MRS observations, attributing the observed infrared excess to a disk-like structure. We applied our CPD models to this dataset and obtained a series of viable fits. The best-fit example is shown in the top panel of Fig.~\ref{fig:6}. The combined atmospheric and disk model successfully reproduces the mid-infrared excess, demonstrating that our parameter grid framework can naturally account for the observed spectral shape without invoking ad hoc emission components.

    YSES~1~b is one of the two Jovian planets in the YSES~1 system at a distance of approximately 94\,pc. Recent atmospheric analyses have yielded somewhat different estimates of its mass and atmospheric properties (e.g., \citealt{Hoch2025, Darcis2026}). \citet{Hoch2025} presented combined JWST/NIRSpec and MIRI/LRS spectra and suggested that the observed infrared excess originates from a circumplanetary disk. By incorporating a modest amount of silicate dust inside cavity, we obtain an excellent match to the observed spectrum of YSES~1~b, as shown in the bottom panel of Fig.~\ref{fig:6}. This result highlights the flexibility of the modeling framework and its ability to accommodate dust-related spectral features when supported by observational evidence. The best-fit model corresponds to a relatively compact dust distribution extending from 31 to 35~$R_{\mathrm{J}}$, which resembles a narrow dust ring rather than a radially extended disk. We note, however, that the cavity interior to this ring is not completely empty. To reproduce the observed silicate emission feature, we additionally include a population of optically thin silicate grains within the cavity. These grains are not part of the main DSHARP dust component discussed below and are therefore not included in the parameter summary table. Such a compact dust distribution may be consistent with a relatively evolved CPD configuration, although the current observations do not uniquely constrain its physical origin.

    To further illustrate the fitting results, we provide in Table~\ref{tab:fitting} the best-fit parameters for the models shown in this section. For YSES~1~b, the optically thin silicate grains introduced to reproduce the silicate feature are not included in the table, which reports only the parameters of the main DSHARP dust component. We explored several disk configurations within the parameter grid, including different disk models. Among these, transitional disk models generally provided a better qualitative match to the observed spectral shape, particularly the infrared excess and overall SED morphology. We emphasize, however, that our fitting procedure is based on a grid exploration rather than a full Bayesian framework. Therefore, the listed parameters should be interpreted as representative best-fit values rather than statistically rigorous estimates with well-defined uncertainties. Nevertheless, the comparison across the model grid allows us to assess qualitatively the robustness of different parameters: parameters that mainly control the overall flux level, such as the disk size or the irradiation level, are typically better constrained, whereas parameters that produce more subtle spectral changes remain more degenerate. These examples illustrate that, despite the simplifying assumptions adopted in the parameter grid, the models are already capable of reproducing key observational signatures seen in current infrared data, demonstrating the potential of the framework as a practical tool for interpreting CPD-related infrared excesses. A more quantitative assessment of parameter uncertainties would require a dedicated statistical analysis, which is beyond the scope of the present work.

    \begin{table*}
        \caption{Representative Best-fit Values}
        \centering
        \begin{tabular}{cccccccccccccc}
            \hline\hline
            \multirow{2}{*}{Source Name} & $T_\mathrm{eff}$ & $R_\mathrm{p}$ & \multirow{2}{*}{$\log g$} & \multirow{2}{*}{Av} & $d$ & $R_\mathrm{in}$ & $R_\mathrm{out}$ & $\Sigma_\mathrm{d,0}$ & \multirow{2}{*}{$\gamma$} & \multirow{2}{*}{$h_\mathrm{d,0}/R_\mathrm{0}$} & \multirow{2}{*}{$\beta$} & $R_\mathrm{cavity}$ & \multirow{2}{*}{$\Sigma_\mathrm{cavity}/\Sigma_\mathrm{d}$} \\
             & [K] & [$R_\mathrm{Jup}$] & & & [pc] & [$R_\mathrm{Jup}$] &
            [$R_\mathrm{Jup}$] & [$\mathrm{g\cdot cm^{-2}}$] & & & & [$R_\mathrm{Jup}$] & \\
            \hline
            GQ~Lup~B & 2713 & 3.6 & 4.93 & 2.42 & 154 & 5 & 55 & 0.5 & $-1.2$ &
            0.07 & 1.1 & 33 & $10^{-4}$
            \\
            \hline
            YSES~1~b & 1740 & 2.5 & 3.5 & 0 & 94 & 5 & 35 & 0.5 & $-0.6$ & 0.09 
            & 1.1 & 31 & $10^{-4}$
            \\
            \hline\hline
        \end{tabular}\\
        \label{tab:fitting}
        \end{table*}

%%%%%%%%%%%%%%%%%%%%%%%%%%%%%%%%%%%%%%%%%%%%%%%%%%%%%%%%%%%%%%
\section{Discussion} \label{sec:discussion}

    Based on the parameter-grid exploration presented in Section~\ref{sec:results}, we have identified several robust trends in the spectral behavior of circumplanetary disks. We find that the overall infrared flux level and spectral slope are primarily governed by the disk geometry and radial extent, while localized substructures such as gaps and cavities introduce wavelength-dependent spectral deficits and excesses through the formation of warm walls. In addition, optically thick and optically thin disks exhibit fundamentally different sensitivities to variations in dust surface density. Pre-transitional disks, in particular, produce intermediate spectral signatures that cannot be trivially mapped to either full or fully transitional disks.

    In this section, we extend the discussion beyond the parameter grid to examine additional physical parameters and modeling assumptions that were not explicitly explored in Section~\ref{sec:results} but may nevertheless have a significant impact on the predicted spectra. These considerations are essential for applying our models to spectral fitting in a more realistic and robust manner.

    \subsection{Planetary Temperature, Radius, and Mass}

        The planetary effective temperature and radius jointly determine the intrinsic luminosity of the central object and therefore strongly influence both the peak wavelength and the overall intensity of the near-infrared spectrum. These parameters are commonly constrained by fitting planetary or brown-dwarf atmospheric models to near-infrared observations, particularly in the $JHK$ bands, in combination with independent estimates of distance and extinction \citep[e.g.,][]{Burrows2006, Allard2012, Spiegel2012, Marley2015}. Such fits typically yield a set of physically plausible parameters, including the effective temperature and radius of the planet. In our models, the disk emission is primarily governed by the irradiation from the central object, i.e., its luminosity. Therefore, a lower effective temperature (for a given radius), such as that inferred for PDS 70 c \citep{Portilla2022, Portilla2023}, would imply a lower luminosity and thus weaker heating of the circumplanetary disk. As a result, the overall flux level would decrease and the spectral energy distribution would shift toward longer wavelengths. While we adopt a fiducial temperature of 2000~K in this work, our results can be qualitatively extended to cooler objects through this scaling with luminosity.

        The planetary mass, by contrast, generally cannot be constrained from spectral energy distributions alone, as it has only a weak and indirect influence on the emergent spectrum at near- and mid-infrared wavelengths. Instead, mass estimates usually rely on dynamical measurements or evolutionary models that combine luminosity and age information \citep[e.g.,][]{Baraffe2003, Marley2007, Spiegel2012}. In this work, the mass does not directly enter the radiative transfer calculations, but may still be relevant when interpreting fitted parameters within an evolutionary context.

        Different atmospheric model families may lead to systematically different best-fit temperatures and radii for the same object. Such differences can propagate into the radiative transfer calculations by altering the radiation field incident on the circumplanetary disk, thereby affecting the predicted disk emission (e.g., \citealp{Rab2019}). As a result, obtaining a satisfactory global fit to combined planetary and disk spectra often requires comparisons across multiple atmospheric model grids rather than reliance on a single model family.

    \subsection{Scattering Effects}\label{sec:scattering}

        \begin{figure}[t!]
            \centering
            \includegraphics[width=0.45\textwidth]{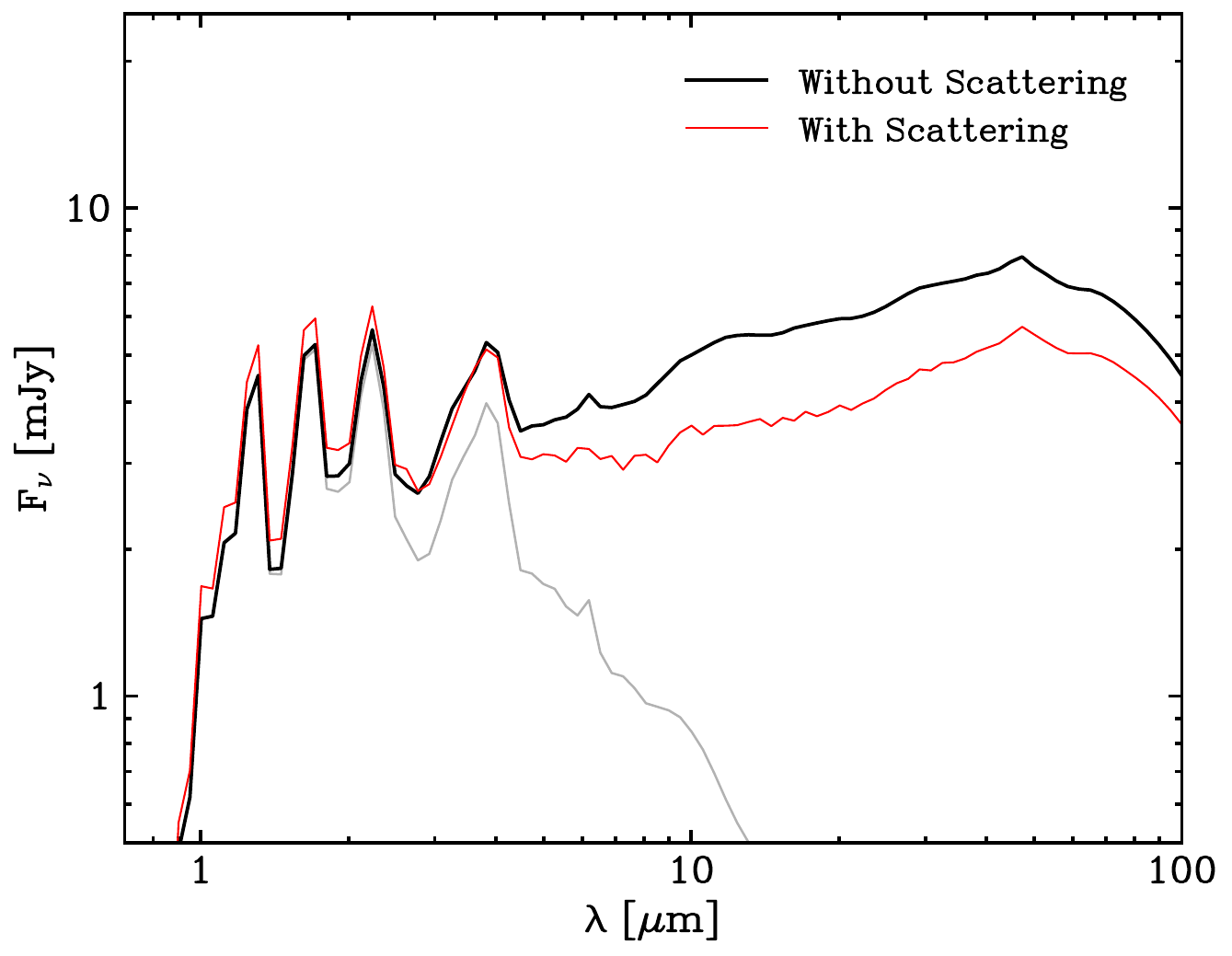}
            \caption{Comparison of the synthetic spectra for the fiducial full disk model with and without dust scattering. The black curve shows the original fiducial model without scattering, while the red curve shows the corresponding model including scattering. The gray curve represents the spectrum of the central atmospheric model.}
            \label{fig:7}
        \end{figure}

        Scattering refers to the process by which dust grains in the disk not only absorb but also scatter a fraction of the radiation emitted by the central planet. In radiative transfer calculations, scattering redistributes photons without strongly altering their wavelengths, such that the scattered light broadly retains the spectral shape of the incident radiation (see, e.g., \citealt{Dullemond2012, Min2016}). As a result, dust in the disk can redirect part of the planetary emission into the observer's line of sight, potentially contributing to the near-infrared flux of the combined system (see, e.g., general discussions of scattering in dusty disks).

        Although this effect is often subtle, it can be important for the interpretation of near-infrared spectra and is sometimes neglected in simplified disk models. If scattering is ignored, a fraction of the near-infrared flux may be incorrectly attributed to the planetary photosphere, potentially leading to an overestimation of the inferred planetary temperature or radius during atmospheric model fitting (e.g., \citealt{Stolker2020, Cugno2024}).

        Furthermore, if scattering is neglected, all incident radiation is effectively assumed to be absorbed by the dust. In reality, a fraction of the radiation may instead be scattered away, reducing the absorbed energy and leading to somewhat lower dust temperatures and thermal emission levels.

        The quantitative impact of scattering depends on several factors, including the dust scattering opacity, grain size distribution, dust density, and the spatial distribution of dust within the disk. To illustrate its potential influence, Fig.~\ref{fig:7} compares the fiducial full disk model with and without scattering. When scattering is included, a modest excess appears at near-infrared wavelengths due to scattered planetary radiation. At the same time, part of the incident radiation is redirected rather than absorbed by the dust, resulting in slightly lower dust temperatures and a corresponding reduction in the mid-infrared thermal emission from the disk.

        In addition, scattering introduces non-local radiative transfer effects, such that the emergent spectrum depends not only on local disk properties but also on the global disk geometry and viewing angle. The magnitude and even the qualitative behavior of scattering signatures can therefore vary significantly between different disk structures and parameter combinations. While the inclusion or omission of scattering does not typically change the broad classification of CPD models (e.g., full, transitional, or evolved disks), it may influence the inferred values of individual disk parameters, particularly when high-quality near-infrared data are available. Because of its strong dependence on poorly constrained dust properties (e.g., albedo and phase function), it remains challenging to robustly assess which disk types would be most affected or to what extent scattering can be used to break degeneracies between different models. In some cases, scattering may provide additional diagnostic power, but it can also introduce new degeneracies that complicate the interpretation of spectral features. Scattering should therefore be taken into account in detailed spectral fitting, especially in studies aiming for quantitative constraints on disk and planetary properties.

    \subsection{Asymmetry and Viewing Angle}

\begin{figure}[t!]
        \centering
        \includegraphics[width=0.45\textwidth]{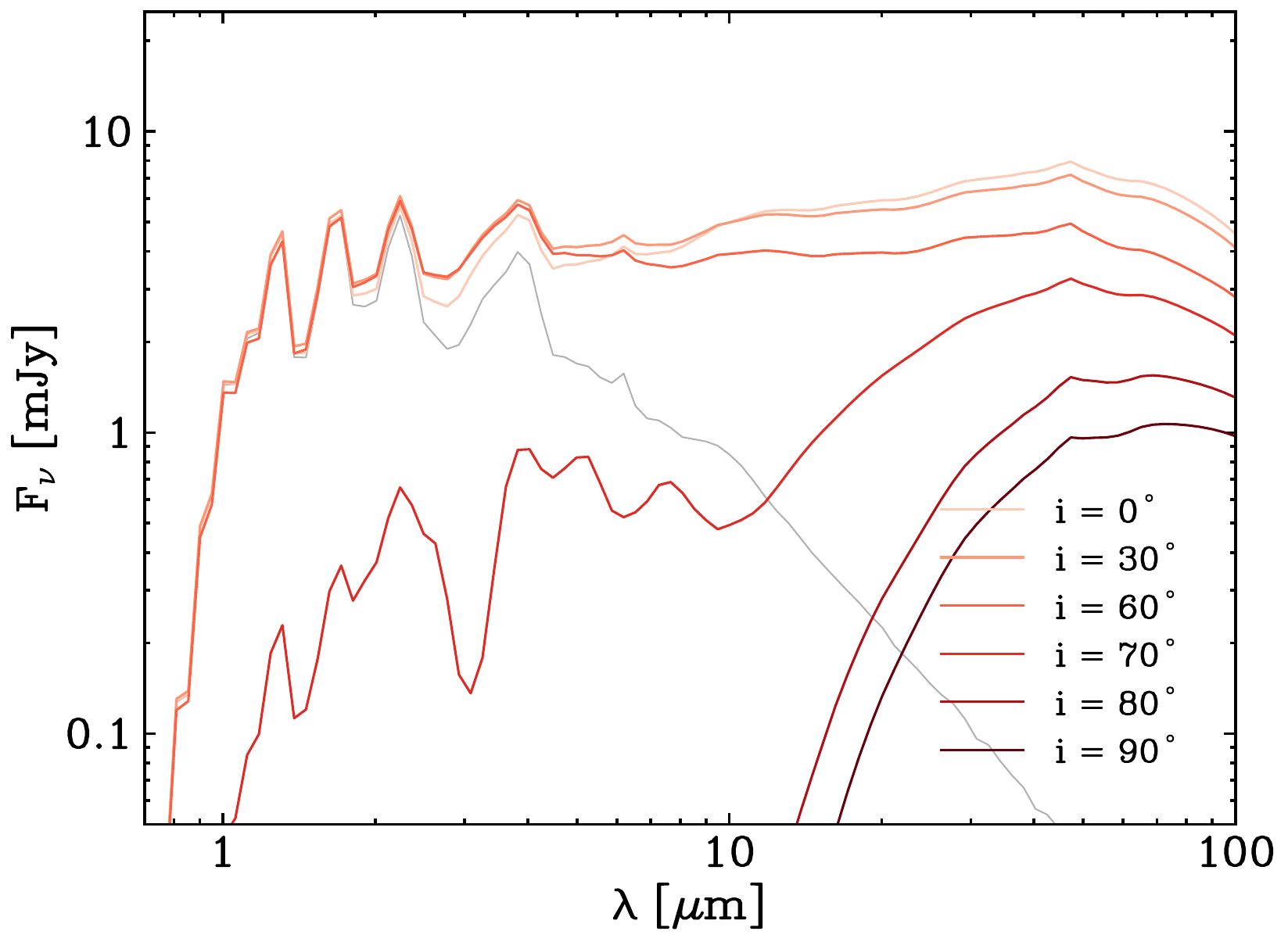}
        \caption{Spectral variations induced by different viewing angles (inclinations) for the same full disk fiducial model. Noticeable differences in the spectral shape and flux level only appear at high inclinations ($\gtrsim 70^\circ$), where obscuration of the central planet becomes significant.}
        \label{fig:8}
    \end{figure}

        Asymmetry and viewing angle are two closely related factors that can influence the observed spectra of CPDs. In the present work, we assume axisymmetric disk structures and therefore do not explicitly account for azimuthal asymmetries. Under this assumption, the dependence of the resulting spectra on the viewing angle is generally weak for moderate inclinations. Significant spectral differences only emerge when the inclination exceeds approximately $70^\circ$, as illustrated in Figure~\ref{fig:8}. At such high inclinations, the central planet becomes increasingly obscured by the disk, making the inferred planetary parameters highly uncertain.

        As a result, spectra obtained at extreme inclinations should be treated with caution and are best considered separately from the main parameter space explored in this study. For the majority of sources observed at lower inclinations, the face-on assumption adopted in our parameter grid is therefore expected to provide a reasonable approximation.

        If disks host asymmetric structures, such as localized overdensities, vortices, or spiral arms, these features can modify the local optical depth and temperature distribution, potentially affecting the emergent spectra \citep[e.g.,][]{Dong2015, Zhu2015_2}. Although such effects are not included in the current grid models, they represent a natural direction for future extensions. For sources with well-characterized asymmetries and available hydrodynamical simulations, incorporating non-axisymmetric structures into radiative transfer modeling may become feasible and, if warranted by the data, should be explored.

    \subsection{Dust Composition and Grain Size Distribution}

        Dust composition and grain size distribution are among the most critical ingredients in determining the spectral properties of circumplanetary disks, as they directly set the wavelength-dependent dust opacity in both absorption and scattering \citep{Draine2003}. These opacities control how efficiently dust grains absorb the planetary radiation, reprocess it into thermal emission, and redistribute photons through scattering, thereby shaping the emergent spectral energy distribution.

        In practice, however, placing robust constraints on dust composition and grain size distribution in CPDs remains highly challenging. Unlike protoplanetary disks, CPDs are typically unresolved and lack multi-wavelength observational coverage, making it difficult to distinguish between different opacity prescriptions \citep[e.g.,][]{Testi2014}. As a result, degeneracies between dust properties and other disk parameters—such as surface density, scale height, or depletion factors—are unavoidable when interpreting spectral data \citep{Birnstiel2012}.

        In this work, we adopt commonly used dust compositions and grain size distributions motivated by standard interstellar and protoplanetary disk models \citep{Pollack1994, Testi2014}. These assumptions provide a reasonable first-order approximation and allow us to systematically explore how disk structure and geometry affect the spectra. While the absolute flux levels and detailed spectral shapes may change under alternative dust prescriptions, the qualitative trends identified through the parameter grid—such as the characteristic spectral signatures of full, evolved, transitional, and pre-transitional disks—are expected to remain robust.
        
        The use of the DSHARP dust opacity mixture tends to smooth out prominent spectral features such as the $10\,\mu$m silicate band, compared to models adopting pure small silicate grains. This explains why strong silicate emission features are not clearly visible in most of our spectra. However, when applying our framework to specific sources (Section~\ref{sec:applications}), different dust compositions—including silicate-rich mixtures—can be readily adopted to reproduce observed spectral features. Since the focus of this work is on the influence of structural parameters rather than detailed dust mineralogy, we defer a systematic exploration of composition-dependent features to future studies.

        Future progress on this front will likely require a combination of improved theoretical modeling and higher-quality observations. In particular, multi-wavelength data spanning the near- to mid-infrared, polarization measurements sensitive to scattering, and constraints from dust evolution models could help break some of the existing degeneracies \citep{Birnstiel2012, Testi2014}. Incorporating more sophisticated dust models, including grain growth, settling, and composition-dependent opacities, will be an important step toward more accurate spectral fitting of CPDs.

    We emphasize that the modeling framework presented in this work is not intended as a direct substitute for formal statistical fitting pipelines. Bayesian fitting frameworks such as DuCKLinG \citep{Woitke2016,Kamp2017,Kaeufer2024} represent a complementary approach that has been successfully applied to protoplanetary disks using parameterized models. The physically motivated CPD grid presented here serves a different purpose, namely to establish systematic links between disk properties and infrared observables, and to provide physical guidance for interpreting disk-related infrared excesses.
    
    Furthermore, at longer wavelengths, particularly in the (sub-)millimeter and radio regime, the emission from circumplanetary disks is expected to be dominated by larger grains, such as millimeter-sized pebbles, which are not included in our current models. These grains can significantly enhance the continuum emission and are often invoked in the context of disk mass estimates and observational detectability with facilities such as ALMA. Therefore, while our models provide useful predictions for infrared observations, they likely underestimate the flux at longer wavelengths. Future work that incorporates grain growth and vertical settling will be necessary to make robust predictions for radio observations and to assess the detectability of circumplanetary disks in this regime.

%%%%%%%%%%%%%%%%%%%%%%%%%%%%%%%%%%%%%%%%%%%%%%%%%%%%%%%%%%%%%%
\section{Conclusions} \label{sec:conclusions}

    In this work, we present an extended modeling framework for circumplanetary disks that combines a broad, physically motivated parameter grid with full radiative transfer simulations performed with \texttt{RADMC-3D}. By varying systematically 12 key disk parameters across a large set of models, and computing their emergent spectra, we construct a self-consistent reference grid that enables a direct connection between CPD structure and near- to mid-infrared observables. This approach allows us to isolate and interpret the spectral impact of individual physical parameters within a controlled and internally consistent framework.

    Beyond the systematic parameter study, we demonstrate the applicability of this framework by fitting representative JWST datasets of planetary-mass companions exhibiting infrared excesses. These examples illustrate that, even under simplified assumptions, the models are capable of reproducing key observational features and provide a physically interpretable explanation for disk-related emission. The framework is therefore well suited as a practical tool for identifying CPD signatures and assessing disk evolutionary stages in current and forthcoming infrared observations.

    Several limitations of the present models should be kept in mind. We assume axisymmetric disk structures, adopt a fixed dust composition and grain size distribution, and neglect scattering in the spectral calculations used for the parameter-grid comparison. Additional physical processes — such as dust growth and settling, non-axisymmetric structures, scattering, inclination effects, and time-dependent accretion — may alter detailed spectral features and will be important extensions in future work.

    Looking ahead, the dense spectral grid presented here, comprising thousands of self-consistently computed radiative transfer models, provides a foundation for more comprehensive CPD studies. We emphasize that the parameter-grid method adopted in this work is intended to complement, rather than replace, formal Bayesian or statistical fitting techniques; it primarily serves to build physical intuition and guide the interpretation of how different disk parameters shape observable spectral features. By combining this framework with complementary observational constraints, such as high-angular-resolution imaging and (sub-)millimeter continuum or line observations, it will be possible to move toward more physically meaningful parameter inference. Ultimately, continued refinement of the model grid — including more realistic dust physics and disk structures — will improve the robustness of CPD interpretation and contribute to a deeper understanding of planet and satellite formation environments.

\section*{Data Availability}

    The model spectra presented in this work are publicly available at 
    \url{https://github.com/MaginaSun/Parameter-grid-spectrum-of-CPDs}. Additional models from the full parameter grid will be made available in future updates of the repository.
    
%%%%%%%%%%%%%%%%%%%%%%%%%%%%%%%%%%%%%%%%%%%%%%%%%%%%%%%%%%%%%%
\begin{acknowledgements}

      We thank the referee for a careful reading of the manuscript and for constructive comments that helped improve the clarity and robustness of this work. We thank Ruobing Dong, Zhizhen Qin, Ruiqi Yang, Pinghui Huang and Gabriele Cugno for many helpful discussions and valuable feedback throughout the development of this project. Their perspectives on disk observations, radiative transfer modeling, and circumplanetary disk physics provided important context that strengthened the robustness of our analysis. We further acknowledge the broader community for open-source tools. G.-D.M.\ acknowledges the support of the Deutsche Forschungsgemeinschaft (DFG) through grant MA~9185/2-1.

\end{acknowledgements}

%%%%%%%%%%%%%%%%%%%%%%%%%%%%%%%%%%%%%%%%%%%%%%%%%%%%%%%%%%%%%%

\bibliographystyle{aa_url}   % aa_url: with links!
\bibliography{ref}

@article{Sun2024,
   author = {Xilei Sun and Pinghui Huang and Ruobing Dong and Shang-Fei Liu},
   doi = {10.3847/1538-4357/ad57c2},
   issn = {0004-637X},
   issue = {1},
   journal = {\apj},
   month = {9},
   pages = {25},
   publisher = {American Astronomical Society},
   title = {Observational Characteristics of Circumplanetary-mass-object Disks in the Era of James Webb Space Telescope},
   volume = {972},
   year = {2024}
}

@ARTICLE{Patapis2025,
       author = {{Patapis}, P. and {Morales-Calder{\'o}n}, M. and {Arabhavi}, A.~M. and {K{\"u}hnle}, H. and {Gasman}, D. and {Cugno}, G. and {Molli{\`e}re}, P. and {Matthews}, E. and {M{\^a}lin}, M. and {Whiteford}, N. and {Lagage}, P.-O. and {Waters}, R. and {Guedel}, M. and {Henning}, Th. and {Vandenbussche}, B. and {Absil}, O. and {Argyriou}, I. and {Barrado}, D. and {Baudoz}, P. and {Boccaletti}, A. and {Bouwman}, J. and {Cossou}, C. and {Coulais}, A. and {Decin}, L. and {Gastaud}, R. and {Glasse}, A. and {Glauser}, A.~M. and {Grant}, S. and {Min}, M. and {Kamp}, I. and {Olofsson}, G. and {Pye}, J. and {Rouan}, D. and {Royer}, P. and {Scheithauer}, S. and {Sun}, X. and {Tremblin}, P. and {Colina}, L. and {Ray}, T.~P. and {{\"O}stlin}, G. and {van Dishoeck}, E.~F. and {Wright}, G.},
        title = "{JWST/MIRI observations of the young TWA 27 system: Hydrocarbon disk chemistry, silicate clouds, and evidence of a circumplanetary disk}",
      journal = {\aap},
         year = 2025,
        month = nov,
       volume = {704},
          eid = {A5},
        pages = {A5},
          doi = {10.1051/0004-6361/202556296},
archivePrefix = {arXiv},
       eprint = {2507.08961},
 primaryClass = {astro-ph.EP},
       adsurl = {https://ui.adsabs.harvard.edu/abs/2025A&A...704A...5P}
}

@article{Cugno2024,
   author = {Gabriele Cugno and Polychronis Patapis and Andrea Banzatti and Michael Meyer and Felix A. Dannert and Tomas Stolker and Ryan J. MacDonald and Klaus M. Pontoppidan},
   doi = {10.3847/2041-8213/ad3cbc},
   issn = {2041-8205},
   issue = {1},
   journal = {\apjl},
   month = {5},
   pages = {L21},
   publisher = {American Astronomical Society},
   title = {Mid-infrared Spectrum of the Disk around the Forming Companion GQ Lup B Revealed by JWST/MIRI},
   volume = {966},
   year = {2024}
}

@article{Hoch2025,
   author = {K. K.W. Hoch and M. Rowland and S. Petrus and E. Nasedkin and C. Ingebretsen and J. Kammerer and M. Perrin and V. D’Orazi and W. O. Balmer and T. Barman and M. Bonnefoy and G. Chauvin and C. Chen and R. J. De Rosa and J. Girard and E. Gonzales and M. Kenworthy and Q. M. Konopacky and B. Macintosh and S. E. Moran and C. V. Morley and P. Palma-Bifani and L. Pueyo and B. Ren and E. Rickman and J. B. Ruffio and C. A. Theissen and K. Ward-Duong and Y. Zhang},
   doi = {10.1038/s41586-025-09174-w},
   issn = {14764687},
   issue = {8073},
   journal = {\nat},
   month = {7},
   pages = {938-942},
   pmid = {40494394},
   publisher = {Nature Research},
   title = {Silicate clouds and a circumplanetary disk in the YSES-1 exoplanet system},
   volume = {643},
   year = {2025}
}

@MISC{Dullemond2012,
       author = {{Dullemond}, C.~P. and {Juhasz}, A. and {Pohl}, A. and {Sereshti}, F. and {Shetty}, R. and {Peters}, T. and {Commercon}, B. and {Flock}, M.},
        title = "{RADMC-3D: A multi-purpose radiative transfer tool}",
 howpublished = {Astrophysics Source Code Library, record ascl:1202.015},
         year = 2012,
        month = feb,
          eid = {ascl:1202.015},
        pages = {ascl:1202.015},
archivePrefix = {ascl},
       eprint = {1202.015},
       adsurl = {https://ui.adsabs.harvard.edu/abs/2012ascl.soft02015D}
}

@article{Birnstiel2018,
   title={The Disk Substructures at High Angular Resolution Project (DSHARP). V. Interpreting ALMA Maps of Protoplanetary Disks in Terms of a Dust Model},
   volume={869},
   ISSN={2041-8213},
   url={http://dx.doi.org/10.3847/2041-8213/aaf743},
   DOI={10.3847/2041-8213/aaf743},
   number={2},
   journal={\apjl},
   publisher={American Astronomical Society},
   author={Birnstiel, Tilman and Dullemond, Cornelis P. and Zhu, Zhaohuan and Andrews, Sean M. and Bai, Xue-Ning and Wilner, David J. and Carpenter, John M. and Huang, Jane and Isella, Andrea and Benisty, Myriam and Pérez, Laura M. and Zhang, Shangjia},
   year={2018},
   month=dec, pages={L45} }

@article{Stolker2020,
   author = {T. Stolker and S. P. Quanz and K. O. Todorov and J. Kühn and P. Mollière and M. R. Meyer and T. Currie and S. Daemgen and B. Lavie},
   doi = {10.1051/0004-6361/201937159},
   issn = {14320746},
   journal = {\aap},
   month = {3},
   publisher = {EDP Sciences},
   title = {MIRACLES: Atmospheric characterization of directly imaged planets and substellar companions at 4-5 μ m: I. Photometric analysis of β Pic b, HIP 65426 b, PZ Tel B, and HD 206893 B},
   volume = {635},
   year = {2020}
}

@article{Blakely2025,
   author = {Dori Blakely and Doug Johnstone and Gabriele Cugno and Anand Sivaramakrishnan and Peter Tuthill and Ruobing Dong and Benjamin J. S. Pope and Loïc Albert and Max Charles and Rachel A. Cooper and Matthew De Furio and Louis Desdoigts and René Doyon and Logan Francis and Alexandra Z. Greenbaum and David Lafreniére and James P. Lloyd and Michael R. Meyer and Laurent Pueyo and Shrishmoy Ray and Joel Sánchez-Bermúdez and Anthony Soulain and Deepashri Thatte and William Thompson and Thomas Vandal},
   doi = {10.3847/1538-3881/ad9b94},
   issn = {0004-6256},
   issue = {3},
   journal = {\aj},
   month = {3},
   pages = {137},
   publisher = {American Astronomical Society},
   title = {The James Webb Interferometer: Space-based Interferometric Detections of PDS 70 b and c at 4.8 μm},
   volume = {169},
   year = {2025}
}

@article{Bae2022,
   title={Molecules with ALMA at Planet-forming Scales (MAPS): A Circumplanetary Disk Candidate in Molecular-line Emission in the AS 209 Disk},
   volume={934},
   ISSN={2041-8213},
   url={http://dx.doi.org/10.3847/2041-8213/ac7fa3},
   DOI={10.3847/2041-8213/ac7fa3},
   number={2},
   journal={\apjl},
   publisher={American Astronomical Society},
   author={Bae, Jaehan and Teague, Richard and Andrews, Sean M. and Benisty, Myriam and Facchini, Stefano and Galloway-Sprietsma, Maria and Loomis, Ryan A. and Aikawa, Yuri and Alarcón, Felipe and Bergin, Edwin and Bergner, Jennifer B. and Booth, Alice S. and Cataldi, Gianni and Cleeves, L. Ilsedore and Czekala, Ian and Guzmán, Viviana V. and Huang, Jane and Ilee, John D. and Kurtovic, Nicolas T. and Law, Charles J. and Gal, Romane Le and Liu, Yao and Long, Feng and Ménard, François and Öberg, Karin I. and Pérez, Laura M. and Qi, Chunhua and Schwarz, Kamber R. and Sierra, Anibal and Walsh, Catherine and Wilner, David J. and Zhang, Ke},
   year={2022},
   month=jul, pages={L20} }

@ARTICLE{Ward2010,
       author = {{Ward}, William R. and {Canup}, Robin M.},
        title = "{Circumplanetary Disk Formation}",
      journal = {\aj},
         year = 2010,
        month = nov,
       volume = {140},
       number = {5},
        pages = {1168-1193},
          doi = {10.1088/0004-6256/140/5/1168},
       adsurl = {https://ui.adsabs.harvard.edu/abs/2010AJ....140.1168W}
}

@ARTICLE{Tanigawa2012,
       author = {{Tanigawa}, Takayuki and {Ohtsuki}, Keiji and {Machida}, Masahiro N.},
        title = "{Distribution of Accreting Gas and Angular Momentum onto Circumplanetary Disks}",
      journal = {\apj},
         year = 2012,
        month = mar,
       volume = {747},
       number = {1},
          eid = {47},
        pages = {47},
          doi = {10.1088/0004-637X/747/1/47},
archivePrefix = {arXiv},
       eprint = {1112.3706},
 primaryClass = {astro-ph.EP},
       adsurl = {https://ui.adsabs.harvard.edu/abs/2012ApJ...747...47T}
}

@ARTICLE{Zhu2015,
       author = {{Zhu}, Zhaohuan},
        title = "{Accreting Circumplanetary Disks: Observational Signatures}",
      journal = {\apj},
         year = 2015,
        month = jan,
       volume = {799},
       number = {1},
          eid = {16},
        pages = {16},
          doi = {10.1088/0004-637X/799/1/16},
archivePrefix = {arXiv},
       eprint = {1408.6554},
 primaryClass = {astro-ph.EP},
       adsurl = {https://ui.adsabs.harvard.edu/abs/2015ApJ...799...16Z}
}

@ARTICLE{Canup2002,
       author = {{Canup}, Robin M. and {Ward}, William R.},
        title = "{Formation of the Galilean Satellites: Conditions of Accretion}",
      journal = {\aj},
         year = 2002,
        month = dec,
       volume = {124},
       number = {6},
        pages = {3404-3423},
          doi = {10.1086/344684},
       adsurl = {https://ui.adsabs.harvard.edu/abs/2002AJ....124.3404C}
}

@ARTICLE{Canup2006,
       author = {{Canup}, Robin M. and {Ward}, William R.},
        title = "{A common mass scaling for satellite systems of gaseous planets}",
      journal = {\nat},
         year = 2006,
        month = jun,
       volume = {441},
       number = {7095},
        pages = {834-839},
          doi = {10.1038/nature04860},
       adsurl = {https://ui.adsabs.harvard.edu/abs/2006Natur.441..834C}
}

@ARTICLE{Szulagyi2017,
       author = {{Szul{\'a}gyi}, J. and {Mayer}, L. and {Quinn}, T.},
        title = "{Circumplanetary discs around young giant planets: a comparison between core-accretion and disc instability}",
      journal = {\mnras},
         year = 2017,
        month = jan,
       volume = {464},
       number = {3},
        pages = {3158-3168},
          doi = {10.1093/mnras/stw2617},
archivePrefix = {arXiv},
       eprint = {1610.01791},
 primaryClass = {astro-ph.EP},
       adsurl = {https://ui.adsabs.harvard.edu/abs/2017MNRAS.464.3158S}
}

@article{Fung2019,
   title={Circumplanetary Disk Dynamics in the Isothermal and Adiabatic Limits},
   volume={887},
   ISSN={1538-4357},
   url={http://dx.doi.org/10.3847/1538-4357/ab53da},
   DOI={10.3847/1538-4357/ab53da},
   number={2},
   journal={\apj},
   publisher={American Astronomical Society},
   author={Fung, Jeffrey and Zhu, Zhaohuan and Chiang, Eugene},
   year={2019},
   month=dec, pages={152} }

@article{Isella2019,
   title={Detection of Continuum Submillimeter Emission Associated with Candidate Protoplanets},
   volume={879},
   ISSN={2041-8213},
   url={http://dx.doi.org/10.3847/2041-8213/ab2a12},
   DOI={10.3847/2041-8213/ab2a12},
   number={2},
   journal={\apjl},
   publisher={American Astronomical Society},
   author={Isella, Andrea and Benisty, Myriam and Teague, Richard and Bae, Jaehan and Keppler, Miriam and Facchini, Stefano and Pérez, Laura},
   year={2019},
   month=jul, pages={L25} }

@article{Benisty2021,
   title={A Circumplanetary Disk around PDS70c},
   volume={916},
   ISSN={2041-8213},
   url={http://dx.doi.org/10.3847/2041-8213/ac0f83},
   DOI={10.3847/2041-8213/ac0f83},
   number={1},
   journal={\apjl},
   publisher={American Astronomical Society},
   author={Benisty, Myriam and Bae, Jaehan and Facchini, Stefano and Keppler, Miriam and Teague, Richard and Isella, Andrea and Kurtovic, Nicolas T. and Pérez, Laura M. and Sierra, Anibal and Andrews, Sean M. and Carpenter, John and Czekala, Ian and Dominik, Carsten and Henning, Thomas and Menard, Francois and Pinilla, Paola and Zurlo, Alice},
   year={2021},
   month=jul, pages={L2} }

@article{Luhman2023,
   title={JWST/NIRSpec Observations of the Planetary Mass Companion TWA 27B*},
   volume={949},
   ISSN={2041-8213},
   url={http://dx.doi.org/10.3847/2041-8213/acd635},
   DOI={10.3847/2041-8213/acd635},
   number={2},
   journal={\apjl},
   publisher={American Astronomical Society},
   author={Luhman, K. L. and Tremblin, P. and Birkmann, S. M. and Manjavacas, E. and Valenti, J. and Alves de Oliveira, C. and Beck, T. L. and Giardino, G. and Lützgendorf, N. and Rauscher, B. J. and Sirianni, M.},
   year={2023},
   month=jun, pages={L36} }

@ARTICLE{Dangelo2008,
       author = {{D'Angelo}, Gennaro and {Lubow}, Stephen H.},
        title = "{Evolution of Migrating Planets Undergoing Gas Accretion}",
      journal = {\apj},
         year = 2008,
        month = sep,
       volume = {685},
       number = {1},
        pages = {560-583},
          doi = {10.1086/590904},
archivePrefix = {arXiv},
       eprint = {0806.1771},
 primaryClass = {astro-ph},
       adsurl = {https://ui.adsabs.harvard.edu/abs/2008ApJ...685..560D}
}

@ARTICLE{Isella2014,
       author = {{Isella}, Andrea and {Chandler}, Claire J. and {Carpenter}, John M. and {P{\'e}rez}, Laura M. and {Ricci}, Luca},
        title = "{Searching for Circumplanetary Disks around LkCa 15}",
      journal = {\apj},
         year = 2014,
        month = jun,
       volume = {788},
       number = {2},
          eid = {129},
        pages = {129},
          doi = {10.1088/0004-637X/788/2/129},
archivePrefix = {arXiv},
       eprint = {1404.5627},
 primaryClass = {astro-ph.EP},
       adsurl = {https://ui.adsabs.harvard.edu/abs/2014ApJ...788..129I}
}

@ARTICLE{Szulagyi2018,
       author = {{Szul{\'a}gyi}, J. and {Plas}, G. van der and {Meyer}, M.~R. and {Pohl}, A. and {Quanz}, S.~P. and {Mayer}, L. and {Daemgen}, S. and {Tamburello}, V.},
        title = "{Observability of forming planets and their circumplanetary discs - I. Parameter study for ALMA}",
      journal = {\mnras},
         year = 2018,
        month = jan,
       volume = {473},
       number = {3},
        pages = {3573-3583},
          doi = {10.1093/mnras/stx2602},
archivePrefix = {arXiv},
       eprint = {1709.04438},
 primaryClass = {astro-ph.EP},
       adsurl = {https://ui.adsabs.harvard.edu/abs/2018MNRAS.473.3573S}
}

@ARTICLE{Burrows2006,
       author = {{Burrows}, A. and {Sudarsky}, D. and {Hubeny}, I.},
        title = "{Theory for the Secondary Eclipse Fluxes, Spectra, Atmospheres, and Light Curves of Transiting Extrasolar Giant Planets}",
      journal = {\apj},
         year = 2006,
        month = oct,
       volume = {650},
       number = {2},
        pages = {1140-1149},
          doi = {10.1086/507269},
archivePrefix = {arXiv},
       eprint = {astro-ph/0607014},
 primaryClass = {astro-ph},
       adsurl = {https://ui.adsabs.harvard.edu/abs/2006ApJ...650.1140B}
}

@ARTICLE{Allard2012,
       author = {{Allard}, F. and {Homeier}, D. and {Freytag}, B.},
        title = "{Models of very-low-mass stars, brown dwarfs and exoplanets}",
      journal = {Philosophical Transactions of the Royal Society of London Series A},
         year = 2012,
        month = jun,
       volume = {370},
       number = {1968},
        pages = {2765-2777},
          doi = {10.1098/rsta.2011.0269},
archivePrefix = {arXiv},
       eprint = {1112.3591},
 primaryClass = {astro-ph.SR},
       adsurl = {https://ui.adsabs.harvard.edu/abs/2012RSPTA.370.2765A}
}

@ARTICLE{Spiegel2012,
       author = {{Spiegel}, David S. and {Burrows}, Adam},
        title = "{Spectral and Photometric Diagnostics of Giant Planet Formation Scenarios}",
      journal = {\apj},
         year = 2012,
        month = feb,
       volume = {745},
       number = {2},
          eid = {174},
        pages = {174},
          doi = {10.1088/0004-637X/745/2/174},
archivePrefix = {arXiv},
       eprint = {1108.5172},
 primaryClass = {astro-ph.EP},
       adsurl = {https://ui.adsabs.harvard.edu/abs/2012ApJ...745..174S}
}

@ARTICLE{Marley2015,
       author = {{Marley}, M.~S. and {Robinson}, T.~D.},
        title = "{On the Cool Side: Modeling the Atmospheres of Brown Dwarfs and Giant Planets}",
      journal = {\araa},
         year = 2015,
        month = aug,
       volume = {53},
        pages = {279-323},
          doi = {10.1146/annurev-astro-082214-122522},
archivePrefix = {arXiv},
       eprint = {1410.6512},
 primaryClass = {astro-ph.EP},
       adsurl = {https://ui.adsabs.harvard.edu/abs/2015ARA&A..53..279M}
}

@ARTICLE{Seifahrt2007,
       author = {{Seifahrt}, A. and {Neuh{\"a}user}, R. and {Hauschildt}, P.~H.},
        title = "{Near-infrared integral-field spectroscopy of the companion to GQ Lupi}",
      journal = {\aap},
         year = 2007,
        month = feb,
       volume = {463},
       number = {1},
        pages = {309-313},
          doi = {10.1051/0004-6361:20066463},
archivePrefix = {arXiv},
       eprint = {astro-ph/0612250},
 primaryClass = {astro-ph},
       adsurl = {https://ui.adsabs.harvard.edu/abs/2007A&A...463..309S}
}

@INPROCEEDINGS{Baraffe2003,
       author = {{Baraffe}, Isabelle and {Chabrier}, Gilles and {Allard}, France and {Hauschildt}, Peter},
        title = "{Evolutionary models for low mass stars and brown dwarfs at young ages}",
    booktitle = {Brown Dwarfs},
         year = 2003,
       editor = {{Mart{\'\i}n}, Eduardo},
       series = {IAU Symposium},
       volume = {211},
        month = jun,
        pages = {41},
       adsurl = {https://ui.adsabs.harvard.edu/abs/2003IAUS..211...41B}
}

@ARTICLE{Marley2007,
       author = {{Marley}, Mark S. and {Fortney}, Jonathan J. and {Hubickyj}, Olenka and {Bodenheimer}, Peter and {Lissauer}, Jack J.},
        title = "{On the Luminosity of Young Jupiters}",
      journal = {\apj},
         year = 2007,
        month = jan,
       volume = {655},
       number = {1},
        pages = {541-549},
          doi = {10.1086/509759},
archivePrefix = {arXiv},
       eprint = {astro-ph/0609739},
 primaryClass = {astro-ph},
       adsurl = {https://ui.adsabs.harvard.edu/abs/2007ApJ...655..541M}
}

@ARTICLE{Min2016,
       author = {{Min}, M. and {Rab}, Ch. and {Woitke}, P. and {Dominik}, C. and {M{\'e}nard}, F.},
        title = "{Multiwavelength optical properties of compact dust aggregates in protoplanetary disks}",
      journal = {\aap},
         year = 2016,
        month = jan,
       volume = {585},
          eid = {A13},
        pages = {A13},
          doi = {10.1051/0004-6361/201526048},
archivePrefix = {arXiv},
       eprint = {1510.05426},
 primaryClass = {astro-ph.EP},
       adsurl = {https://ui.adsabs.harvard.edu/abs/2016A&A...585A..13M}
}

@ARTICLE{Dong2015,
       author = {{Dong}, Ruobing and {Hall}, Cassandra and {Rice}, Ken and {Chiang}, Eugene},
        title = "{Spiral Arms in Gravitationally Unstable Protoplanetary Disks as Imaged in Scattered Light}",
      journal = {\apjl},
         year = 2015,
        month = oct,
       volume = {812},
       number = {2},
          eid = {L32},
        pages = {L32},
          doi = {10.1088/2041-8205/812/2/L32},
archivePrefix = {arXiv},
       eprint = {1510.00396},
 primaryClass = {astro-ph.SR},
       adsurl = {https://ui.adsabs.harvard.edu/abs/2015ApJ...812L..32D}
}

@ARTICLE{Zhu2015_2,
       author = {{Zhu}, Zhaohuan and {Dong}, Ruobing and {Stone}, James M. and {Rafikov}, Roman R.},
        title = "{The Structure of Spiral Shocks Excited by Planetary-mass Companions}",
      journal = {\apj},
         year = 2015,
        month = nov,
       volume = {813},
       number = {2},
          eid = {88},
        pages = {88},
          doi = {10.1088/0004-637X/813/2/88},
archivePrefix = {arXiv},
       eprint = {1507.03599},
 primaryClass = {astro-ph.SR},
       adsurl = {https://ui.adsabs.harvard.edu/abs/2015ApJ...813...88Z}
}

@INPROCEEDINGS{Schneeberger2024,
       author = {{Schneeberger}, Antoine and {Mousis}, Olivier and {Lunine}, Jonathan},
        title = "{Impact of Jupiter's heating and self-shadowing on the structure of its circumplanetary disk}",
    booktitle = {European Geosciences Union General Assembly 2024 (EGU24)},
         year = 2024,
       series = {EGU General Assembly Conference Abstracts},
        month = apr,
          eid = {8573},
        pages = {8573},
          doi = {10.5194/egusphere-egu24-8573},
       adsurl = {https://ui.adsabs.harvard.edu/abs/2024EGUGA..26.8573S}
}

@ARTICLE{Sagynbayeva2025,
       author = {{Sagynbayeva}, Sabina and {Li}, Rixin and {Kuznetsova}, Aleksandra and {Zhu}, Zhaohuan and {Jiang}, Yan-Fei and {Armitage}, Philip J.},
        title = "{Circumplanetary Disks are Rare Around Planets at Large Orbital Radii: A Parameter Survey of Flow Morphology Around Giant Planets}",
      journal = {\apj},
         year = 2025,
        month = jul,
       volume = {987},
       number = {2},
          eid = {216},
        pages = {216},
          doi = {10.3847/1538-4357/add934},
archivePrefix = {arXiv},
       eprint = {2410.14896},
 primaryClass = {astro-ph.EP},
       adsurl = {https://ui.adsabs.harvard.edu/abs/2025ApJ...987..216S}
}

@ARTICLE{Krapp2024,
       author = {{Krapp}, Leonardo and {Kratter}, Kaitlin M. and {Youdin}, Andrew N. and {Ben{\'\i}tez-Llambay}, Pablo and {Masset}, Fr{\'e}d{\'e}ric and {Armitage}, Philip J.},
        title = "{A Thermodynamic Criterion for the Formation of Circumplanetary Disks}",
      journal = {\apj},
         year = 2024,
        month = oct,
       volume = {973},
       number = {2},
          eid = {153},
        pages = {153},
          doi = {10.3847/1538-4357/ad644a},
archivePrefix = {arXiv},
       eprint = {2402.14638},
 primaryClass = {astro-ph.EP},
       adsurl = {https://ui.adsabs.harvard.edu/abs/2024ApJ...973..153K}
}

@ARTICLE{Schulik2025,
       author = {{Schulik}, Matth{\"a}us and {Bitsch}, Bertram and {Johansen}, Anders and {Lambrechts}, Michiel},
        title = "{The influence of dust growth on the observational properties of circumplanetary discs}",
      journal = {\aap},
         year = 2025,
        month = mar,
       volume = {695},
          eid = {A126},
        pages = {A126},
          doi = {10.1051/0004-6361/202038068},
archivePrefix = {arXiv},
       eprint = {2501.15521},
 primaryClass = {astro-ph.EP},
       adsurl = {https://ui.adsabs.harvard.edu/abs/2025A&A...695A.126S}
}

@ARTICLE{Portilla2022,
       author = {{Portilla-Revelo}, B. and {Kamp}, I. and {Rab}, Ch. and {van Dishoeck}, E.~F. and {Keppler}, M. and {Min}, M. and {Muro-Arena}, G.~A.},
        title = "{Self-consistent modelling of the dust component in protoplanetary and circumplanetary disks: the case of PDS 70}",
      journal = {\aap},
         year = 2022,
        month = feb,
       volume = {658},
          eid = {A89},
        pages = {A89},
          doi = {10.1051/0004-6361/202141764},
archivePrefix = {arXiv},
       eprint = {2111.08648},
 primaryClass = {astro-ph.EP},
       adsurl = {https://ui.adsabs.harvard.edu/abs/2022A&A...658A..89P}
}

@ARTICLE{Chen2022,
       author = {{Chen}, Xueqing and {Szul{\'a}gyi}, Judit},
        title = "{Observability of forming planets and their circumplanetary discs - IV. With JWST and ELT}",
      journal = {\mnras},
         year = 2022,
        month = oct,
       volume = {516},
       number = {1},
        pages = {506-528},
          doi = {10.1093/mnras/stac1976},
archivePrefix = {arXiv},
       eprint = {2112.12821},
 primaryClass = {astro-ph.EP},
       adsurl = {https://ui.adsabs.harvard.edu/abs/2022MNRAS.516..506C}
}

@ARTICLE{Portilla2023,
       author = {{Portilla-Revelo}, B. and {Kamp}, I. and {Facchini}, S. and {van Dishoeck}, E.~F. and {Law}, C. and {Rab}, Ch. and {Bae}, J. and {Benisty}, M. and {{\"O}berg}, K. and {Teague}, R.},
        title = "{Constraining the gas distribution in the PDS 70 disc as a method to assess the effect of planet-disc interactions}",
      journal = {\aap},
         year = 2023,
        month = sep,
       volume = {677},
          eid = {A76},
        pages = {A76},
          doi = {10.1051/0004-6361/202346607},
archivePrefix = {arXiv},
       eprint = {2306.16850},
 primaryClass = {astro-ph.EP},
       adsurl = {https://ui.adsabs.harvard.edu/abs/2023A&A...677A..76P}
}

@ARTICLE{Taylor2025,
       author = {{Taylor}, Aster G. and {Adams}, Fred C.},
        title = "{Radiative signatures of circumplanetary disks and envelopes during the late stages of giant planet formation}",
      journal = {\icarus},
         year = 2025,
        month = jan,
       volume = {425},
          eid = {116327},
        pages = {116327},
          doi = {10.1016/j.icarus.2024.116327},
archivePrefix = {arXiv},
       eprint = {2409.12733},
 primaryClass = {astro-ph.EP},
       adsurl = {https://ui.adsabs.harvard.edu/abs/2025Icar..42516327T}
}

@ARTICLE{Draine2003,
       author = {{Draine}, B.~T.},
        title = "{Scattering by Interstellar Dust Grains. I. Optical and Ultraviolet}",
      journal = {\apj},
         year = 2003,
        month = dec,
       volume = {598},
       number = {2},
        pages = {1017-1025},
          doi = {10.1086/379118},
archivePrefix = {arXiv},
       eprint = {astro-ph/0304060},
 primaryClass = {astro-ph},
       adsurl = {https://ui.adsabs.harvard.edu/abs/2003ApJ...598.1017D}
}

@ARTICLE{Birnstiel2012,
       author = {{Birnstiel}, T. and {Andrews}, S.~M. and {Ercolano}, B.},
        title = "{Can grain growth explain transition disks?}",
      journal = {\aap},
         year = 2012,
        month = aug,
       volume = {544},
          eid = {A79},
        pages = {A79},
          doi = {10.1051/0004-6361/201219262},
archivePrefix = {arXiv},
       eprint = {1206.5802},
 primaryClass = {astro-ph.SR},
       adsurl = {https://ui.adsabs.harvard.edu/abs/2012A&A...544A..79B}
}

@INPROCEEDINGS{Testi2014,
       author = {{Testi}, L. and {Birnstiel}, T. and {Ricci}, L. and {Andrews}, S. and {Blum}, J. and {Carpenter}, J. and {Dominik}, C. and {Isella}, A. and {Natta}, A. and {Williams}, J.~P. and {Wilner}, D.~J.},
        title = "{Dust Evolution in Protoplanetary Disks}",
    booktitle = {Protostars and Planets VI},
         year = 2014,
       editor = {{Beuther}, Henrik and {Klessen}, Ralf S. and {Dullemond}, Cornelis P. and {Henning}, Thomas},
        month = jan,
        pages = {339-361},
          doi = {10.2458/azu_uapress_9780816531240-ch015},
archivePrefix = {arXiv},
       eprint = {1402.1354},
 primaryClass = {astro-ph.SR},
       adsurl = {https://ui.adsabs.harvard.edu/abs/2014prpl.conf..339T}
}

@ARTICLE{Pollack1994,
       author = {{Pollack}, James B. and {Hollenbach}, David and {Beckwith}, Steven and {Simonelli}, Damon P. and {Roush}, Ted and {Fong}, Wesley},
        title = "{Composition and Radiative Properties of Grains in Molecular Clouds and Accretion Disks}",
      journal = {\apj},
         year = 1994,
        month = feb,
       volume = {421},
        pages = {615},
          doi = {10.1086/173677},
       adsurl = {https://ui.adsabs.harvard.edu/abs/1994ApJ...421..615P}
}

@ARTICLE{Kaeufer2024,
       author = {{Kaeufer}, T. and {Min}, M. and {Woitke}, P. and {Kamp}, I. and {Arabhavi}, A.~M.},
        title = "{Bayesian analysis of the molecular emission and dust continuum of protoplanetary disks}",
      journal = {\aap},
         year = 2024,
        month = jul,
       volume = {687},
          eid = {A209},
        pages = {A209},
          doi = {10.1051/0004-6361/202449936},
archivePrefix = {arXiv},
       eprint = {2405.06486},
 primaryClass = {astro-ph.EP},
       adsurl = {https://ui.adsabs.harvard.edu/abs/2024A&A...687A.209K}
}

@ARTICLE{Batygin2025,
       author = {{Batygin}, Konstantin and {Adams}, Fred C.},
        title = "{Determination of Jupiter's primordial physical state}",
      journal = {Nature Astronomy},
         year = 2025,
        month = jun,
       volume = {9},
        pages = {835-844},
          doi = {10.1038/s41550-025-02512-y},
archivePrefix = {arXiv},
       eprint = {2505.12652},
 primaryClass = {astro-ph.EP},
       adsurl = {https://ui.adsabs.harvard.edu/abs/2025NatAs...9..835B}
}

@ARTICLE{Rab2019,
   author = {{Rab}, C. and {Kamp}, I. and {Ginski}, C. and {Oberg}, N. and
        {Muro-Arena}, G.~A. and {Dominik}, C. and {Waters}, L.~B.~F.~M. and
        {Thi}, W.-F. and {Woitke}, P.},
    title = "{Observing the gas component of circumplanetary disks around wide-orbit planet-mass companions in the (sub)mm regime}",
  journal = {\aap},
archivePrefix = "arXiv",
   eprint = {1902.04096},
 primaryClass = "astro-ph.SR",
     year = 2019,
    month = apr,
   volume = 624,
      eid = {A16},
    pages = {A16},
      doi = {10.1051/0004-6361/201834899},
   adsurl = {https://ui.adsabs.harvard.edu/abs/2019A%26A...624A..16R}
}

@ARTICLE{Dullemond2001,
       author = {{Dullemond}, C.~P. and {Dominik}, C. and {Natta}, A.},
        title = "{Passive Irradiated Circumstellar Disks with an Inner Hole}",
      journal = {\apj},
         year = 2001,
        month = oct,
       volume = {560},
       number = {2},
        pages = {957-969},
          doi = {10.1086/323057},
archivePrefix = {arXiv},
       eprint = {astro-ph/0106470},
 primaryClass = {astro-ph},
       adsurl = {https://ui.adsabs.harvard.edu/abs/2001ApJ...560..957D}
}

@ARTICLE{Pinte2006,
       author = {{Pinte}, C. and {M{\'e}nard}, F. and {Duch{\^e}ne}, G. and {Bastien}, P.},
        title = "{Monte Carlo radiative transfer in protoplanetary disks}",
      journal = {\aap},
         year = 2006,
        month = dec,
       volume = {459},
       number = {3},
        pages = {797-804},
          doi = {10.1051/0004-6361:20053275},
archivePrefix = {arXiv},
       eprint = {astro-ph/0606550},
 primaryClass = {astro-ph},
       adsurl = {https://ui.adsabs.harvard.edu/abs/2006A&A...459..797P}
}

@INPROCEEDINGS{Espaillat2014,
       author = {{Espaillat}, C. and {Muzerolle}, J. and {Najita}, J. and {Andrews}, S. and {Zhu}, Z. and {Calvet}, N. and {Kraus}, S. and {Hashimoto}, J. and {Kraus}, A. and {D'Alessio}, P.},
        title = "{An Observational Perspective of Transitional Disks}",
    booktitle = {Protostars and Planets VI},
         year = 2014,
       editor = {{Beuther}, Henrik and {Klessen}, Ralf S. and {Dullemond}, Cornelis P. and {Henning}, Thomas},
        month = jan,
        pages = {497-520},
          doi = {10.2458/azu_uapress_9780816531240-ch022},
archivePrefix = {arXiv},
       eprint = {1402.7103},
 primaryClass = {astro-ph.SR},
       adsurl = {https://ui.adsabs.harvard.edu/abs/2014prpl.conf..497E}
}

@ARTICLE{Whitney2003a,
       author = {{Whitney}, Barbara A. and {Wood}, Kenneth and {Bjorkman}, J.~E. and {Wolff}, Michael J.},
        title = "{Two-dimensional Radiative Transfer in Protostellar Envelopes. I. Effects of Geometry on Class I Sources}",
      journal = {\apj},
         year = 2003,
        month = jul,
       volume = {591},
       number = {2},
        pages = {1049-1063},
          doi = {10.1086/375415},
archivePrefix = {arXiv},
       eprint = {astro-ph/0303479},
 primaryClass = {astro-ph},
       adsurl = {https://ui.adsabs.harvard.edu/abs/2003ApJ...591.1049W}
}

@ARTICLE{Whitney2023b,
       author = {{Whitney}, Barbara A. and {Wood}, Kenneth and {Bjorkman}, J.~E. and {Cohen}, Martin},
        title = "{Two-dimensional Radiative Transfer in Protostellar Envelopes. II. An Evolutionary Sequence}",
      journal = {\apj},
         year = 2003,
        month = dec,
       volume = {598},
       number = {2},
        pages = {1079-1099},
          doi = {10.1086/379068},
archivePrefix = {arXiv},
       eprint = {astro-ph/0309007},
 primaryClass = {astro-ph},
       adsurl = {https://ui.adsabs.harvard.edu/abs/2003ApJ...598.1079W}
}

@ARTICLE{Whitney2004,
       author = {{Whitney}, Barbara. A. and {Indebetouw}, R{\'e}my and {Bjorkman}, J.~E. and {Wood}, Kenneth},
        title = "{Two-Dimensional Radiative Transfer in Protostellar Envelopes. III. Effects of Stellar Temperature}",
      journal = {\apj},
         year = 2004,
        month = dec,
       volume = {617},
       number = {2},
        pages = {1177-1190},
          doi = {10.1086/425608},
       adsurl = {https://ui.adsabs.harvard.edu/abs/2004ApJ...617.1177W}
}

@ARTICLE{Woitke2016,
       author = {{Woitke}, P. and {Min}, M. and {Pinte}, C. and {Thi}, W.-F. and {Kamp}, I. and {Rab}, C. and {Anthonioz}, F. and {Antonellini}, S. and {Baldovin-Saavedra}, C. and {Carmona}, A. and {Dominik}, C. and {Dionatos}, O. and {Greaves}, J. and {G{\"u}del}, M. and {Ilee}, J.~D. and {Liebhart}, A. and {M{\'e}nard}, F. and {Rigon}, L. and {Waters}, L.~B.~F.~M. and {Aresu}, G. and {Meijerink}, R. and {Spaans}, M.},
        title = "{Consistent dust and gas models for protoplanetary disks. I. Disk shape, dust settling, opacities, and PAHs}",
      journal = {\aap},
         year = 2016,
        month = feb,
       volume = {586},
          eid = {A103},
        pages = {A103},
          doi = {10.1051/0004-6361/201526538},
archivePrefix = {arXiv},
       eprint = {1511.03431},
 primaryClass = {astro-ph.EP},
       adsurl = {https://ui.adsabs.harvard.edu/abs/2016A&A...586A.103W}
}

@ARTICLE{Kamp2017,
       author = {{Kamp}, I. and {Thi}, W.-F. and {Woitke}, P. and {Rab}, C. and {Bouma}, S. and {M{\'e}nard}, F.},
        title = "{Consistent dust and gas models for protoplanetary disks. II. Chemical networks and rates}",
      journal = {\aap},
         year = 2017,
        month = nov,
       volume = {607},
          eid = {A41},
        pages = {A41},
          doi = {10.1051/0004-6361/201730388},
archivePrefix = {arXiv},
       eprint = {1707.07281},
 primaryClass = {astro-ph.SR},
       adsurl = {https://ui.adsabs.harvard.edu/abs/2017A&A...607A..41K}
}

@ARTICLE{Darcis2026,
       author = {{Darcis}, Michiel and {Haffert}, Sebastiaan Y. and {Stolker}, Tomas and {van Capelleveen}, Richelle F. and {Kenworthy}, Matthew A. and {de Visser}, Pieter J. and {Close}, Laird M. and {Guyon}, Olivier and {Hedglen}, Alexander D. and {Johnson}, Parker T. and {Kautz}, Maggie Y. and {Kueny}, Jay K. and {Li}, Jialin and {Long}, Joseph D. and {Lumbres}, Jennifer and {Males}, Jared R. and {McEwen}, Eden A. and {McLeod}, Avalon L. and {Pearce}, Logan A. and {Schatz}, Lauren and {Van Gorkom}, Kyle},
        title = "{The spectral energy distribution of YSES 1 b and its circumplanetary disc}",
      journal = {arXiv e-prints},
         year = 2026,
        month = may,
          eid = {arXiv:2605.26805},
        pages = {arXiv:2605.26805},
          doi = {10.48550/arXiv.2605.26805},
archivePrefix = {arXiv},
       eprint = {2605.26805},
 primaryClass = {astro-ph.EP},
       adsurl = {https://ui.adsabs.harvard.edu/abs/2026arXiv260526805D}
}

\end{document}